\documentclass[aps,prd,11pt,reprint,nofootinbib,showkeys,superscriptaddress]{revtex4-1} 
\usepackage{amsmath,amssymb,amsthm,mathrsfs,bbm}
\usepackage{latexsym,amscd,amsbsy,amsfonts,dsfont}
\usepackage{graphicx}
\usepackage[utf8]{inputenc}
\usepackage{marvosym}
\usepackage{float}
\usepackage{slashed}
\usepackage{cancel}
\usepackage{multirow}
\usepackage{soul}
\usepackage{physics}
\usepackage{comment}
\usepackage[normalem]{ulem}
\usepackage[usenames,dvipsnames]{xcolor}
\usepackage[
      colorlinks=true,
      linktocpage=true,
      breaklinks=true,
      linkcolor=Aquamarine,
      urlcolor=Purple,
      filecolor=black,
      citecolor=Purple,
      menucolor=black,
      pdfstartview=FitV,
      bookmarksopen=true
      ]{hyperref}
\usepackage{tikz}
\usepackage[usenames,dvipsnames]{xcolor}

\makeindex

\begin{document}

\title{Classical and quantum field theory in a box with moving boundaries:\\ A numerical study of the Dynamical Casimir Effect}
\author{Alberto Garc\'{i}a Mart\'{i}n-Caro}
\email{agmcaro@gmail.com}
\affiliation{Department of Physics, University of the Basque Country UPV/EHU, 48940, Bilbao, Spain}
\author{Gerardo Garc\'ia-Moreno}
\email{ggarcia@iaa.es}
\affiliation{Instituto de Astrof\'{\i}sica de Andaluc\'{\i}a (IAA-CSIC), Glorieta de la Astronom\'{\i}a, 18008 Granada, Spain}
\author{Javier Olmedo}
\email{javolmedo@ugr.es}
\affiliation{Departamento de F\'isica Te\'orica y del Cosmos, Universidad de Granada, Granada-18071, Spain}
\author{Jose M. S\'anchez Vel\'azquez}
\affiliation{Instituto de F\'isica Te\'orica UAM/CSIC, c/ Nicol\'as Cabrera 13-15, Cantoblanco, 28049, Madrid, Spain}
\email{jm.sanchez.velazquez@csic.es}

\begin{abstract}
We present a detailed description of a quantum scalar field theory within a flat spacetime confined to a cavity with perfectly reflecting moving boundaries. Moreover, we establish an equivalence between this time-dependent setting and a field theory on an acoustic metric with static Dirichlet boundary conditions. We discuss the classical and quantum aspects of the theory from the latter perspective, accompanied by the introduction of novel numerical techniques designed for the (nonperturbative) computation of particle production attributed to the Dynamical Casimir effect, applicable to arbitrary boundary trajectories. As an illustrative example of these methodologies, we compute the particle production for a massless field in 1+1 dimensions. Notably, our approaches readily extend to encompass scenarios involving massive fields and higher dimensions. 
\end{abstract}


\maketitle

\section{Introduction}
\label{Sec:Introduction}

The study of linear field theories in flat spacetimes with time-dependent boundary conditions has been a subject of research for several decades and finds applications in various fields. One of the most popular models involves an accelerated mirror asymptotically approaching the speed of light~\cite{Birrell1982}, leading to particle creation while the mirror is accelerating. This phenomenon results in the formation of a thermal state in the asymptotic limit. Moreover, there are variations of this model, considering imperfect reflectors \cite{Obadia:2001hj,Haro:2007jr}, alternative mirror trajectories \cite{Su:2016gnj}, leading to applications in entanglement harvesting \cite{Cong:2018vqx}, and some of them that are even relevant for holography \cite{Reyes2021,Akal:2022qei,Kumar2023}. 

The situation where the field is confined in a cavity has also received considerable attention. If the boundaries remain at rest, one can study the well-known static Casimir effect \cite{casimir}. It has served, for instance,  as an example for the study of a nontrivial renormalized stress-energy tensor in quantum field theory on flat spacetimes. The more general case in which one or both boundaries follow nontrivial time-dependent trajectories has also received considerable attention, leading to the so-called dynamical Casimir effect (DCE). The simplest case under study has been a massless (1+1)-dimensional scalar field theory \cite{Moore1970,Castagnino:1984ej,Dalvit:1998qs,PhysRevA.74.013806}. It has also been generalized to 3+1 dimensions~\cite{Crocce:2001zz}, and for non-flat scenarios \cite{Celeri:2008ui,Lock:2016rmg} (see Refs. \cite{Barbado:2018qod,Barbado:2021wnn} for an extension to more general curved spacetimes). These scenarios are also a suitable and interesting arena for the study of relativistic quantum information~\cite{Friis:2012tb,Alsing_2012,Bruschi:2012pd,Lindkvist:2013dha,Sabin:2015pwa,Romualdo:2019eur,Bosco:2019ayk,DelGrosso:2020rbi}. For a recent review on the topic, we refer the reader to \cite{Dodonov:2020eto}. It is interesting to note that DCE has been recently proposed as a phenomenon that can be studied experimentally using coplanar wave guides ended in superconducting quantum interference devices (SQUIDS)~\cite{PhysRevA.73.063818,PhysRevA.82.052509,Doukas:2014bja,Pasi2013} {as well as in semiconductor sheets irradiated by a pulsed laser \cite{Agnesi_2008,Naylor1}. 

Following the latter proposal we have recently shown~\cite{GarciaMartin-Caro:2023jjq} that some configurations of the boundaries (within current experimental capabilities) can mimic an accelerating mirror, giving rise to nearly thermal particle production in some frequency bands.  With this in mind, our aim in this manuscript is twofold. On one hand, we show that a (1+1)-dimensional field theory with moving cavities in a flat metric is equivalent to a (1+1)-dimensional field theory with static cavities on a time-dependent inhomogeneous metric. But more importantly, the resulting (1+1)-dimensional spacetime line element is actually an acoustic metric, of the kind considered in analogue black hole models~\cite{Barcelo2005} for the study of Hawking radiation in the laboratory. On the other hand, we provide a refined formalism for the study of the classical and quantum dynamics of this field theory, with special interest in probing the non-perturbative regime, namely, without the assumption that the boundaries move with small amplitudes, velocities and/or accelerations. We report here the numerical tools that we developed for the study of these configurations in~\cite{GarciaMartin-Caro:2023jjq}, which constitute the basis for the study of other nonperturbative regimes in several interesting models already considered in the literature, see e. g.~\cite{Bosco:2019ayk} and references therein. 

Concretely, this manuscript is organized as follows. In Sec.~\ref{Sec:classical} we describe the equivalence between a (1+1)-dimensional field theory in Minkowski with moving boundaries and a (1+1)-dimenonal field theory in an acoustic metric with static boundaries. In Sec.~\ref{Sec:Classical_Theory} we provide the canonical formulation of the classical theory. We then provide a detailed quantization of the theory in Sec.~\ref{Sec:Quantization}. Sec.~\ref{Sec:Solution} introduces the numerical tools used to compute the particle production. We also provide some examples where we apply them. We conclude with Sec.~\ref{Sec:conclusions}. At the end, the reader can find two appendixes. The first of them contains a description of the conformal transformation method which is an alternative method that we have used to contrast our results~\ref{Subsec:Conformal_Transformation_Method}, and the second one discusses the massive scalar field theory in 3D~\ref{Subsec:3+1}.

\emph{Notation and conventions:} We choose natural units in which $ \hbar = c = 1$, we use the convention of mostly plus signs for the signature of the metric and we introduce the bar notation denoting complex conjugation, e.g. $\bar{z}$ represents the complex conjugate of $z$.

\section{Equivalence between moving boundaries in Minkowski and acoustic metrics}
\label{Sec:classical}
The system that we are going to study is a free real scalar field $\phi$ in the $(1+1)$-dimensional Minkowski spacetime. Its action is given as usual by 
\begin{align}\label{eq:action}
    S = - \frac{1}{2} \int d^2 x \sqrt{- \eta} \eta^{\mu \nu} \partial_{\mu} \phi \partial_{\nu} \phi, 
\end{align}
where $\eta_{\mu \nu}$ represents the flat spacetime metric in an arbitrary coordinate system. A straightforward variation of the action leads to the Klein-Gordon equation  
\begin{align}
    \frac{1}{\sqrt{-\eta}} \partial_{\mu} \left( \sqrt{-\eta} \eta^{\mu \nu} \partial_{\nu} \right) \phi = 0. 
\end{align}
We are interested in imposing time-dependent perfectly reflecting boundary conditions, i.e., the problem of a scalar field confined within the region bounded by the trajectories of two moving ``mirrors" (we will use interchangeably the word mirror and boundaries where perfectly reflecting boundary conditions are imposed). To be more explicit, we want to study the evolution of the field for arbitrary times $t \in (- \infty, \infty)$, $t$ being an inertial time coordinate in Minkowski spacetime, for the region bounded between to trajectories $f(t)$ and $g(t)$ such that the distance among the two boundaries is always greater than zero $L(t) = g(t) - f(t)$. This means that the field $\phi(t,x)$ will be subjected to the Dirichlet boundary conditions $\phi(t,f(t)) = \phi(t,g(t))= 0 $. As a final remark, we will limit our discussion to trajectories of the boundaries that are at rest at early times and become at rest also at late times. This ensures that we have a well-defined and unique notion of in and out vacua.

We will illustrate now how the boundary conditions trivialize (i.e. they become time-independent) for a particular choice of coordinates. In those coordinates, the metric acquires the form of an acoustic metric.
Let us consider the transformation 
\begin{equation}\label{eq:coordtrans}
\tau=t\quad {\rm and}\quad \xi=L_0\frac{x-f(t)}{L(t)},
\end{equation}
where $L_0$ is a dimensionful constant that we choose to represent the distance of the plates at some asymptotic past time where the plates are relatively at rest. In the following, we  make $L_0=1$ and take it as a reference length scale. The flat spacetime metric $ds^2=-dt^2+dx^2$ now becomes
\begin{equation}\label{eq:accoustic-g}
\begin{split}
ds^2=-\left(1-V^2(\tau,\xi)\right)d\tau^2+2L(\tau)V(\tau,\xi)d\tau d\xi+L^2(\tau)d\xi^2,
\end{split}
\end{equation}
where
\begin{equation}
V(\tau,\xi)=\xi \dot L(\tau)+\dot f(\tau).
\end{equation}
Here, the dot will represent differentiation with respect to the time coordinate $\tau$. The line element in Eq.~\eqref{eq:accoustic-g} has the form of an acoustic metric (see for instance Ref. \cite{Barcelo2005}), with $V(\tau,\xi)$ playing the role of the velocity of an effective fluid. 

It is interesting to ask if this metric has acoustic horizons, i.e. points at which $V(\tau,\xi)=\pm 1$. The answer is in the negative. For instance, if $\dot L(\tau)=0$, the horizon corresponds to $\dot f(\tau)=\pm 1$, which is not allowed since this implies that the mirrors move at the speed of light. On the other hand, if $\dot f(\tau)=0$, then $\xi \dot L(\tau)=\pm 1$. Since $\dot f(\tau)=0$, it implies $\dot L(\tau)\in(-1,1)$. Together with $\xi\in[0,1]$, the condition $\xi \dot L(\tau)=\pm 1$ will not be satisfied. Finally, if both $\dot f(\tau)\neq 0$ and $\dot L(\tau)\neq 0$, we have 
\begin{equation}
    \xi \dot L(\tau)+\dot f(\tau)=\pm 1.
\end{equation}
In this case, let us recall that $L(\tau)=g(\tau)-f(\tau)$ and hence $\dot L(\tau)\in(-2,2)$, since $\dot g(\tau)\in(-1,1)$ and $\dot f(\tau)\in(-1,1)$. Then, since we also have $\xi\in[0,1]$, we conclude that
\begin{equation}
    0< \frac{\pm 1-\dot f(\tau)}{\dot g(\tau)-\dot f(\tau)}<1,
\end{equation}
but it is not difficult to see that there are no choices of $\dot g(\tau)\in(-1,1)$ and $\dot f(\tau)\in(-1,1)$ fulfilling the inequality. Hence, no horizons will form. However, there are settings in which it is possible to engineer an effective superluminal motion of the boundaries. In such setup, it seems that acoustic horizons form. This setup is convoluted and rich enough to deserve a dedicated study.  

\section{Classical canonical formulation}\label{Sec:Classical_Theory}
This section is devoted to the analysis of the classical theory. Our study departs from more conventional analysis since we will adopt a Hamiltonian formulation since the theory is amenable to be easily quantized. We will also introduce the set of {\it in} and {\it out} modes that will be relevant later for the discussion of particle production. We will consider configurations that are stationary in the asymptotic past and future. Hence, our results will be free of the typical ambiguities one finds in general time-dependent settings as we advanced before.  

We start with an Arnowitt-Deser-Misner decomposition of the $(1+1)$-dimensional spacetime metric~\cite{adm}. The spacelike ``hypersurface'' defined by $\tau=\textrm{const.}$, yields the lapse, shift and spatial sections metric
\begin{equation}
N=1,\quad N_{\xi}=\xi L \dot L+L \dot f, \quad h_{\xi\xi}=L^2,
\end{equation}
respectively. Then, the conjugate momentum to the scalar field  is related to the field derivatives by
\begin{equation}
\pi_\phi := L \dot{\phi} - \frac{1}{L} \left(\xi \dot{L} + \dot{f} \right) \phi'.
\label{Eq:CanonicalMomentum}
\end{equation}

Implementing the Legendre transformation, the original action is written in terms of the field variable and the momenta:
\begin{equation}
S=\int d \tau \left[\left(\int_{0}^{1} d \xi \,\dot \phi \pi_\phi\right)-H_{T}\right],
\end{equation}
with the Hamiltonian being defined as
\begin{equation}
H_{T}=\int_{0}^{1} d \xi\left[\frac{N}{2} \frac{\pi_{\phi}^{2}}{L}+\frac{N}{2 L}\left(\phi^{\prime}\right)^{2}+\left(\xi \frac{\dot L}{L}+\frac{\dot f}{L}\right) \phi^{\prime} \pi_{\phi}\right].
\label{Eq:Total_Hamiltonian}
\end{equation}
Here, the prime denotes differentiation with respect to the spatial coordinate $\xi$. The canonical Poisson brackets at equal times are
\begin{equation}\label{eq:position-poisson}
\left\{ \phi( \tau,\xi), \pi_{\phi}(\tau,\tilde{\xi})\right\}=\delta(\xi-\tilde{\xi}).
\end{equation}
For our purposes it is convenient to introduce the following Fourier decomposition in phase space

\begin{align}\label{eq:fourier}
 & \phi(\tau,\xi)=  \sum_{n=1}^{\infty} \phi_n (\tau) \sin (n \pi \xi), \nonumber \\
 & \pi_\phi(\tau,\xi)=  \sum_{n=1}^{\infty} \pi_n (\tau) \sin (n \pi \xi). 
\end{align}
which guarantees that the phase space variables fulfill the boundary conditions, namely, $\phi(\xi=0)=\phi(\xi=1)=0$. This expansion for the field $\phi(\tau, \xi)$ is such that it ensures that the boundary conditions are obeyed $\phi(\tau, 0) = \phi(\tau, 1) = 0 $. The momentum $\pi_{\phi}(\tau,\xi)$ does not vanish on the boundaries when they are in motion due to the second term in Eq.~\eqref{Eq:CanonicalMomentum}. However, it is still convenient to use the $\left\{\sin(n\pi \xi)\right\}$ base to expand it. As long as we restrict ourselves to observables away from the boundary, the expansion is perfectly suited for every computation. A thorough treatment of the boundary requires to treat carefully the boundary terms in the action and it is not within the scope of this paper, see e.g.~\cite{Calogeracos1995} for a discussion.

In terms of the Fourier modes, the action is given by
\begin{equation}
 S=\int\,d \tau \left[ \sum_{n=1}^{\infty}\frac{1}{2} \pi_{n} \dot\phi_{n}-H_T\right],
\end{equation}
where now the Hamiltonian is also expressed in terms of the Fourier amplitudes and modes as
\begin{align}\nonumber
 &H_T=\sum_{n=1}^{\infty}\frac{N}{4 L}\left[\pi_{n}^{2}+(n \pi)^{2} \phi_{n}^{2}\right]-\frac{\dot L}{4 L} \pi_{n} \phi_{n}\\\nonumber
 &+\sum_{m}\left(1-\delta_{n m}\right)\frac{n m}{n^{2}-m^{2}}
 \\
 &\times\left[\frac{\dot f}{L}\left((-1)^{n+m}-1\right)+\frac{\dot L}{L}(-1)^{n+m}\right]  \pi_{m} \phi_{n}.
\end{align}
See~\cite{PhysRevA.49.433,PhysRevA.57.2311} for similar analysis. Here, the modes satisfy the Poisson brackets structure
\begin{equation}\label{eq:fourier-poisson}
\left\{\phi_{n}, \pi_{m} \right\}=2 \delta_{n m}.
\end{equation}
The Hamilton equations of motion encoding the dynamics of the field, derived within the previous framework are 
\begin{align}
& \dot \phi_{n}=\frac{N}{L} \pi_{n}-\frac{\dot L}{2 L} \phi_{n} +2 \sum_{m}\left(1-\delta_{m n}\right)\frac{m n}{m^{2}-n^{2}}\nonumber \\\label{eq:dotphi1d}
& \times\left[\frac{\dot f}{L}\left((-1)^{m+n}-1\right)+\frac{\dot L}{L}(-1)^{m+n}\right]  \phi_{m},\\
&\dot{\pi}_{n}=-\frac{N}{L}(n \pi)^{2} \phi_{n}+\frac{\dot L}{2 L} \pi_{n} +2 \sum_{m}\left(1-\delta_{n m}\right)\frac{n m}{m^{2}-n^{2}}\nonumber \\\label{eq:dotpi1d}
& \times\left[\frac{\dot f}{L}\left((-1)^{n+m}-1\right)+\frac{\dot L}{L}(-1)^{n+m}\right]  \pi_{m}.
\end{align}
From these equations of motion it becomes obvious that if $\dot L(\tau)\neq 0$ and/or $\dot f(\tau)\neq 0$ there will be dynamical mode mixing. Its origin can be traced back to Eq.~\eqref{Eq:Total_Hamiltonian}, where the $\phi^{\prime} \pi_{\phi}$ coupling ensures that this mode-mixing emerges. In that sense, this process is closer in spirit to particle production in inhomogeneous cosmological spacetimes. 

For many practical purposes, specially regarding the quantization of the classical theory but also from the classical viewpoint itself, it is useful to complexify the phase space. At the end of the day, when one computes observables, one only needs to ensure that the quantities of interest are real. In our case, although we will be working with the Fourier modes as if they were complex, they must satisfy the reality conditions
\begin{equation}\label{eq:realcond}
    \bar\phi_n(\tau) = \phi_n(\tau),\quad \bar\pi_n(\tau) = \pi_n(\tau).
\end{equation}
With this in mind, we can compute complex solutions to the equations of motion \eqref{eq:dotphi1d} and \eqref{eq:dotpi1d}, given suitable initial data. Any solution can be written as a vector in phase space with an infinite number of components (a pair configuration and momenta for each mode). We will denote a solution ${\bf U}(\tau)$ as
\begin{equation}
    {\bf U}(\tau) =\Big(\phi_1(\tau),\pi_1(\tau), \phi_2(\tau),\pi_2(\tau), \cdots\Big),
\end{equation}
and we will call $\cal{S}^{\mathbb{C}}$ the complexified space of solutions, i.e. $\boldsymbol{U} (\tau) \in \cal{S}^{\mathbb{C}}$. Our space of solutions is endowed with a natural Klein-Gordon product which is preserved under the evolution, i.e. a map $ \text{KG}: \cal{S}^{\mathbb{C}} \times \cal{S}^{\mathbb{C}} \rightarrow \mathbb{C}$. Given two complex solutions ${\bf U}^{(1)}(\tau)$ and ${\bf U}^{(2)}(\tau)$, this product is expressed as
\begin{align}\nonumber
&\langle {\bf U}^{(1)}(\tau), {\bf U}^{(2)}(\tau)\rangle=i \\ \nonumber
&\times\int_{0}^{1} d \xi \,\left(\bar{\phi}^{(1)}(\tau, \xi) \pi_{\phi}^{(2)}(\tau, \xi)-\bar{\pi}^{(1)}(\tau, \xi) \phi^{(2)}(\tau, \xi)\right)\\ \label{eq:inner-prod}
&=\frac{i}{2} \sum_{n=1}^{\infty} \bar{\phi}_{n}^{(1)}(\tau) \pi_{n}^{(2)}(\tau)-\bar{\pi}_{n}^{(1)}(\tau) \phi_{n}^{(2)}(\tau),
\end{align}
and since it is time independent, it is unaffected by the choice of $\tau$ at which we decide to evaluate it. However, we notice that this product is not positive definite and hence it cannot be used straightforwardly to endow our complexified space of solutions with an actual inner product. The standard procedure to do it is to choose a subspace of the complex phase space where the Klein-Gordon product~\eqref{eq:inner-prod} is positive definite. This subspace ${\cal S}^+ \subset {\cal S}^{\mathbb{C}}$ is usually called the positive frequency sector of the theory, since in standard it is spanned by positive frequency plane waves (i.e. modes whose time dependence with respect to an inertial time coordinate are $ \sim e^{ -i \omega t}$). The complementary sector of the phase space where the inner product is negative corresponds to the complex conjugate solutions of the positive frequency one, the sector $ \sim e^{ i \omega t}$. We can endow such complementary space with an inner product straightforwardly also simply by taking the opposite of the Klein-Gordon product for such solutions. This space ${\cal S}^-$ turns out to be the complex conjugate space to ${\cal S}^+$. It can be shown that this construction~\cite{Wald1995} based on the Klein-Gordon product automatically gives rise to a decomposition of the complexified space of solutions as a direct sum of two orthogonal subspaces:
\begin{align}
    \cal{S}^{\mathbb{C}} = {\cal S}^+ \oplus {\cal S}^-
\end{align}
endowed with the inner product introduced above.\footnote{This decomposition into positive and negative frequency sectors is equivalent to choosing a complex structure, namely, a linear symplectomorphism whose square is minus the identity, which combined with the symplectic structure, provides an inner product in that phase space.}   

Given this decomposition, we will now choose a basis of (orthonormal) complex solutions. In general, we will denote this basis of solutions as $\left({\bf u}^{(I)}, \bar{{\bf u}}^{(I)}\right)$, with $I=1,2,\ldots$ This basis satisfies the following identities
\begin{equation}\label{eq:basis}
    \langle {\bf u}^{(I)}, {\bf u}^{\left(J\right)}\rangle=\delta^{I J}, \;\langle {\bf u}^{(I)}, \bar{{\bf u}}^{\left(J\right)}\rangle=0, \;\langle\bar{{\bf u}}^{(I)}, \bar{{\bf u}}^{\left(J\right)}\rangle=-\delta^{I J}.
\end{equation}
Any solution to the equations of motion can be expressed as a linear combination of the elements of the basis, but since we are interested on real solutions, they must be of the form
\begin{equation}\label{eq:sol}
{\bf U}(\tau)=\sum_{I=1}^\infty a_I {\bf u}^{(I)}(\tau)+\bar a_I \bar{{\bf u}}^{(I)}(\tau),
\end{equation}
where $\bar a_I$ and $a_I$ are the creation and annihilation variables. 
In components, if we introduce the following notation, the solutions will be
\begin{equation}\label{eq:sol-mode}
U_{n+\epsilon}(\tau)=\sum_{I=1}^\infty a_I u_{n+\epsilon}^{(I)}(\tau)+\bar a_I \bar u_{n+\epsilon}^{(I)}(\tau),
\end{equation}
with $\epsilon=0,1$, and $n$ an integer that labels the corresponding Fourier mode. Then, if $\epsilon=0$ we have $U_{n}(\tau)=\phi_n(\tau)$, while for $\epsilon=1$ we get $U_{n+1}(\tau)=\pi_n(\tau)$.

Note that either combination \eqref{eq:sol} or \eqref{eq:sol-mode} guaranties that $\bar U_n(\tau)=U_n(\tau)$, for all $n$, in agreement with Eq. \eqref{eq:realcond}. 

It is interesting to note that the creation and annihilation variables can be defined as
\begin{equation}
a_{I}=\left\langle {\bf u}^{(I)}(\tau), {\bf U}(\tau)\right\rangle, \quad \bar{a}_{I}=-\left\langle\bar{{\bf u}}^{(I)}(\tau), {\bf U}(\tau)\right\rangle.
\end{equation}
One can see that the Poisson algebra~\eqref{eq:fourier-poisson} together with the relations in Eq.~\eqref{eq:basis} imply
\begin{equation}\label{eq:amu-poisson}
\left\{a_{I}, \bar{a}_{J}\right\}=-i\delta^{IJ},\quad \left\{a_{I}, {a}_{J}\right\}=0=\left\{\bar a_{I}, \bar{a}_{J}\right\}.
\end{equation}
Let us express the Poisson algebra in~\eqref{eq:fourier-poisson} as 
\begin{equation}
\left\{U_{n}(\tau), U_{m}(\tau)\right\}=2\Omega_{nm},
\end{equation}
being $\Omega_{nm}$ the elements of the block matrix 
\begin{equation}
\boldsymbol{\Omega} = \lim_{N\to\infty}\bigoplus_{n=1}^N\left(\begin{array}{cc}
0 & 1 \\
-1 & 0
\end{array}\right)
\end{equation}
Notice that the Poisson algebra~\eqref{eq:amu-poisson} implies the closure conditions \cite{Agullo:2020uii}
\begin{equation}\label{eq:closure-cond}
\frac{i}{2} \sum_{I=1}^{\infty}\left(-u_{n}^{(I)}(\tau) \bar u_{m}^{(I)}(\tau)+\bar{u}_{n}^{(I)}(\tau) u_{m}^{(I)}(\tau)\right)  = \Omega_{nm}.
\end{equation}
Finally, given two basis with elements ${\bf u}^{(I)}(\tau)$ and ${\bf w}^{(I)}(\tau)$, respectively, they will be related by a Bogoliubov transformation
\begin{equation}\label{eq:bogou}
    {\bf u}^{(I)}(\tau) = \sum_{J=1}^{\infty} \alpha_{IJ}{\bf w}^{(J)}(\tau)+\beta_{IJ}\bar {\bf w}^{(J)}(\tau).
\end{equation}
where $\alpha_{IJ}$ and $\beta_{IJ}$ are the so-called Bogoliubov coefficients. They can be expressed in terms of the components of the two solutions as
\begin{align}\nonumber
&\alpha_{IJ} = \langle {\bf w}^{(J)}(\tau), {\bf u}^{(I)}(\tau)\rangle=\\\nonumber
&\frac{i}{2} \sum_{n=1}^{\infty} {}^{w}\bar{\phi}_{n}^{(J)}(\tau) {}^{u}\pi_{n}^{(I)}(\tau)-{}^{w}\bar{\pi}_{n}^{(J)}(\tau) {}^{u}\phi_{n}^{(I)}(\tau),\\
&\beta_{IJ} = -\langle \bar {\bf w}^{(J)}(\tau), {\bf u}^{(I)}(\tau)\rangle=\\ \nonumber
&-\frac{i}{2} \sum_{n=1}^{\infty} {}^{w}{\phi}_{n}^{(J)}(\tau) {}^{u}\pi_{n}^{(I)}(\tau)-{}^{w}{\pi}_{n}^{(J)}(\tau) {}^{u}\phi_{n}^{(I)}(\tau).
\end{align}
Conversely, the inverse relation to Eq. \eqref{eq:bogou} is given by 
\begin{equation}\label{eq:bogow}
    {\bf w}^{(I)}(\tau) = \sum_{J=1}^{\infty} \bar\alpha_{JI}{\bf u}^{(J)}(\tau)-\beta_{JI}\bar {\bf u}^{(J)}(\tau).
\end{equation}
They satisfy several conditions. From the normalization conditions of the basis $\left({\bf u}^{(I)}, \bar{{\bf u}}^{(I)}\right)$, we obtain
\begin{align}\label{eq:bogo1}
&\sum_{K=1}^{\infty} \alpha_{I K} \bar{\alpha}_{J K}-\beta_{I K} \bar{\beta}_{J K}=\delta_{I J}, \\\label{eq:bogo2}
&\sum_{K=1}^{\infty} \alpha_{I K} \beta_{J K}-\beta_{I K} \alpha_{J K}=0 .
\end{align}
From the normalization conditions of the basis $\left({\bf w}^{(I)}, \bar{{\bf w}}^{(I)}\right)$, we get
\begin{align}\label{eq:bogo3}
&\sum_{K=1}^{\infty} \bar\alpha_{K I} {\alpha}_{KJ}-\beta_{KI} \bar{\beta}_{ KJ}=\delta_{I J}, \\\label{eq:bogo4}
&\sum_{K=1}^{\infty} \bar\alpha_{KI} \beta_{ K J}-\beta_{KI} \bar\alpha_{K J}=0.
\end{align}
Moreover, if $\bar b_I$ and $b_I$ are coefficients of the expansion of a given solution in another basis, that we represent as ${\bf w}^{(I)}(\tau)$, then
\begin{equation}\label{eq:bogob2a}
    b_{I} = \sum_{J=1}^{\infty} \alpha_{JI}a_{J}+\bar\beta_{JI}\bar a_{J},
\end{equation}
or equivalently,
\begin{equation}\label{eq:inv-bogo}
    a_{I} = \sum_{J=1}^{\infty} \bar\alpha_{IJ}b_{J}-\bar\beta_{IJ}\bar b_{J}.
\end{equation}
In summary, the field $\phi(\tau,\xi)$ and its momentum $\pi(\tau,\xi)$ at any time can be written as in Eq. \eqref{eq:fourier}, but replacing the Fourier modes $\phi_n$ and $\pi_n$ by
\begin{equation}\label{eq:mode-aadagg}
\begin{split}
 &\phi_n(\tau)=  \sum_{I=1}^\infty a_I u_{2n-1}^{(I)}(\tau)+\bar a_I \bar u_{2n-1}^{(I)}(\tau),\\ &\pi_n(\tau)=\sum_{I=1}^\infty a_I u_{2n}^{(I)}(\tau)+\bar a_I \bar u_{2n}^{(I)}(\tau),
 \end{split}
\end{equation}
As a final remark, let us introduce some special set of modes specially suited for the kind of trajectories that we will consider in this work. First of all, since we assume that the plates are stationary in the past, before a given time $\tau_i$ at which the motion of the boundaries begins, there is a natural basis (the $in$ basis) of orthonormal complex solutions given by the solutions to the equations with initial conditions 
\begin{align}\nonumber
{\bf u}^{(1)}\left(\tau_{0}\right)&=\left(\frac{1}{\sqrt{\omega_{1}}},-i \sqrt{\omega_{1}}, 0, 0, \cdots\right),\\\nonumber
{\bf u}^{(2)}\left(\tau_{0}\right)&=\left(0,0,\frac{1}{\sqrt{\omega_{2}}},-i \sqrt{\omega_{2}}, 0, 0, \cdots\right),\\\nonumber
&\vdots\\
{\bf u}^{(I)}\left(\tau_{0}\right)&=\left(0,0, \ldots, \frac{1}{\sqrt{\omega_{I}}}, -i \sqrt{\omega_{I}},0, 0, \ldots\right), \label{Eq:initial_modes}\\ \nonumber
&\vdots
\end{align}
and their complex conjugate, where $\omega_{1}=\pi$, $\omega_{2}=2\pi$, $\ldots$ are frequencies of the modes $n=1,2,\ldots$, respectively. Notice that $\tau_0 < \tau_i$ as our working assumption. This set of modes corresponds to standard plane waves in the past region where the boundaries remain at rest. We will use them later in order to perform a standard quantization of the theory as we do in flat spacetime.

In the far future, the plates become at rest again at some given time $\tau_f$. Hence, we can also introduce a second basis of modes representing plane waves in the far future region. These modes are called {\it out} modes, and they are those defined by the initial conditions that have the same functional form as in Eq.~\eqref{Eq:initial_modes}. However, notice that now these initial conditions are given at times after the motion has stopped $\tau_0 > \tau_f$. Although it might seem on first sight that these modes are equal to the {\it in} modes once we evolve them throughout all the spacetime and obtain their explicit $\tau$-dependence, it is not the case due to the non-trivial evolution that the modes undergo in the time interval $\tau \in (\tau_i, \tau_f)$. 

\section{Quantum theory}
\label{Sec:Quantization}
This section is devoted to the quantization of the classical theory presented in the previous section. We will first argue how it is possible to associate a Fock space (and hence a vacuum) with a given set of modes, and how observables can be computed in such framework, providing explicit expressions in terms of the classical modes for some of the most relevant observables. We will illustrate how to compute the particle creation due to the moving boundaries. In fact, we will write down explicit expressions for all the relevant observables (for different states of interest) in terms of the Bogoliubov coefficients $\alpha$ and $\beta$ relating the {\it in} and {\it out} modes.

The quantization of a classical system is tantamount to representing the classical canonical algebra~\eqref{eq:position-poisson} as an algebra of operators acting on a Hilbert space. Actually, this procedure allows us to represent only linear operators of the theory. To represent non-linear operators (evaluated at the coincident spacetime points) we need to deal with renormalization issues. Although we will consider product of linear operators, i.e. $n$-point functions, we will avoid the coincidence limit.

In the case at hand we want to represent the classical algebra associated with the real scalar field; namely,
\begin{equation}\label{eq:position-poisson-2}
\left[\hat\phi(\tau,\xi), \hat \pi_{\phi}(\tau,\tilde{\xi})\right]=i \delta(\xi-\tilde{\xi})\hat{\bf I},
\end{equation}
with $\hat{\bf I}$ the identity operator. Considering the Fourier expansion in Eq. \eqref{eq:fourier}, the above commutation relations yield
\begin{equation}\label{eq:qp-commut}
\left[\hat\phi_{n}(\tau), \hat\pi_{m}(\tau) \right]=2 i\delta_{n m}\hat{\bf I}.
\end{equation}
Hence, if we encode the field and momentum operators in the vector $\hat {\bf U}(\tau)$, and given any orthonormal basis of positive frequency complex solutions $\left({\bf u}^{(I)}, \bar{\bf u}^{(I)}\right)$, the creation and annihilation operators are easily obtained from
\begin{equation}
\hat a_{I}=\left\langle {\bf u}^{(I)}(\tau), \hat {\bf U}(\tau)\right\rangle, \quad 
\hat{a}^{\dagger}_{I}=-\left\langle\bar{\bf u}^{(I)}(\tau), \hat {\bf U}(\tau)\right\rangle.
\end{equation}
One can then straightforwardly seen that the commutation relations \eqref{eq:qp-commut} imply 
\begin{equation}\label{eq:amu-commut}
\left[\hat a_{I}, \hat a^{\dagger}_{J}\right]=\delta_{IJ}\hat{\bf I},\quad \left[\hat a_{I}, \hat {a}_{J}\right]=0=\left[a^{\dagger}_{I}, a^{\dagger}_{J}\right],
\end{equation}
for all $I,J$. The vacuum state is then defined as the one which is annihilated by all the annihilation operators, namely,
\begin{equation}
\hat a_{I}|0\rangle = 0,\quad I=1,2,\ldots.
\end{equation}
Interestingly, in this canonical framework, we can express $\hat\phi(\tau,\xi)$ and $\hat\pi_\phi(\tau,\xi)$ in terms of $\hat a_{I}$ and $\hat a^{\dagger}_{I}$ by means of the classical expressions \eqref{eq:fourier} and \eqref{eq:mode-aadagg}, just promoting the annihilation and creation variables to quantum operators. 

The basic observables in the quantum theory are the $n$-point functions as we have already advanced. They are the expectation values on a state $\hat \rho$ of products of configuration and momentum variables evaluated at different positions and times. Actually, using the Hamilton equations that relates configuration and momenta, one can obtain any of them by taking time and spatial derivatives of the $n$-point function
\begin{equation}\label{eq:n-point}
G_\rho(\tau_1,\xi_1;\ldots;\tau_n,\xi_n)={\rm Tr}\left(\hat\rho\,\hat \phi(\tau_1,\xi_1)\cdots \hat \phi(\tau_n,\xi_n)\right).
\end{equation}
At the end of the day, after Fourier expansion, this $n$-point function can be computed out of expectation values of products of the creation and annihilation variables on the state $\hat\rho$. Below, we will discuss some particular states of special interest. In all cases, we focus our attention on boundaries at rest in the asymptotic past and future. Hence, the set of modes introduced at the end of the previous section can be used to define two different quantizations and hence two different vacua, they are called the {\it in} and {\it out} vacua. The former is the vacuum that an inertial observer at early times, before the motion of the boundaries starts, would probe. The second is the one associated with inertial observers at late times, i.e. after the motion of the boundaries stops. Hence, our main aim is to choose the {\it in}-quantization, as well as a suitable state there (which could be the vacuum or not) and compare it with the {\it out}-quantization, i.e., with the {\it out}-vacuum. This will tell us the amount of particles in the final state. We will denote the {\it in}-vacuum as $\ket{0_{\text{in}}}$ and the {\it out}-vacuum as $\ket{0_{\text{out}}}$. They will be annihilated by the $\hat a_I$ and the $\hat b_{I}$ operators, respectively, i.e. $\hat a_I\ket{0_{\text{in}}} = 0$ and $\hat b_I\ket{0_{\text{out}}} = 0$. The relation among the operators $a$ and $b$ is given by Eqs.~\eqref{eq:bogob2a} and~\eqref{eq:inv-bogo}. 

Among the set of possible states that we could consider, we will focus on the so-called Gaussian states. These states are entirely characterized by their $1$ and $2$-point functions~\cite{Adesso2014,Serafini2017}. They can be regarded as Gaussian probability distributions. These states display many interesting properties, like for instance, the preservation of a Gaussian state under time evolution if the Hamiltonian is, at most, quadratic in the creation and annihilation operators (as it is the case for us). Furthermore, many of the most relevant states belong to this category; vacuum, squeezed, coherent and thermal states. 

We will now compute the most relevant observables that capture particle production due to the motion of the boundaries once they become at rest in the future. We will take any of the previous Gaussian states as initial state. The two independent quadratic observables that are more relevant for our purposes are: 
\begin{equation}
    \hat O^{(1)}_{IJ}=\hat b_I\hat b_J,\quad  \hat O^{(2)}_{IJ}=\hat b^\dagger_I\hat b_J,
\end{equation}
with their expectation values computed as 
\begin{equation}
    \expval{\hat O^{(1)}_{IJ}} = {\rm Tr} \left[ \hat \rho \,  \hat b_I \hat b_J \right], \quad \expval{\hat O^{(2)}_{IJ}} = {\rm Tr} \left[ \hat \rho \, \hat b^{\dagger}_I \hat b_J \right],
    \label{Eq:Basic_Operators}
\end{equation}
in some state $\hat \rho$. All the other quadratic operators are related to these ones through hermitian conjugation or directly from the canonical commutation relations. We also notice that knowing these expectation values is equivalent to knowing the $2$-point function as introduced in~\eqref{eq:n-point}. The diagonal part of $\expval{\hat O^{(2)}_{II}}$ (no sum over I) is specially relevant since it corresponds to the number of produced particles on the mode labeled by $I$.  
\subsection{Vacuum state}
For linear field theories in the vacuum state $\hat \rho_{\textrm{vac}}=|0_{\text{in}}\rangle\langle 0_{\text{in}} |$, the moving boundaries create particles in pairs. Hence, only even powers of products of annihilation and creation variables will contribute to the $n$-point functions. As an example, we compute the simplest non-trivial expectation values, encapsulated in Eq.~\eqref{Eq:Basic_Operators}. One can easily verify that they are given by 
\begin{equation}
\ev{\hat  O^{(1)}_{IJ}}_{\textrm{vac}}=\sum_{K=1}^{\infty}\alpha_{KI}\bar\beta_{KJ},
\end{equation}
\begin{equation}
\ev{\hat  O^{(2)}_{IJ}}_{\textrm{vac}} =
\sum_{K=1}^{\infty} \beta_{KI}\bar\beta_{KJ}.
\end{equation}
The diagonal part of this expectation value, namely $I=J$, can be identified with the number of particles in the mode $I$, as we have already advanced. It acquires the standard expression for particle production: a sum of the squared modulus of the $\beta$-coefficients 
\begin{equation}
\ev{\hat N_I}_{\textrm{vac}} =\sum_{K=1}^{\infty}|\beta_{IK}|^2.
\label{Eq:number_vacuum}
\end{equation}

\subsection{Thermal state}
Another important family of states are those of thermal equilibrium at a finite temperature $T$ (in natural units). For these states the particle production due to the DCE is known to be enhanced~\cite{Crocce:2001zz}. However, the enhancement is basically because the DCE acts as an amplifier, in the sense that if there is an already existing number of particles, the final number of particles grows linearly with the initial number. For thermal states, which are states that lack quantum correlations and behave classically, the particle creation departs from the spontaneous particle creation in vacuum. In that sense, it is hard to determine whether the resulting number of particles has a purely quantum origin. Let us particularize this discussion to our setup.

Consider that, for $\tau<\tau_i$, both mirrors are at rest and we prepare the field in a thermal state at a temperature $T$. As the distance between the mirrors is finite, the mode spectrum will be discrete and we will be able to describe such state in terms of a density matrix of the Gibbs type, i.e.,
\begin{equation}\label{eq:thermal-state}
    \hat \rho_T=Z^{-1}\exp{-\hat H/T}
\end{equation}
with $Z={\rm Tr}\left[\exp{- \hat H/T}\right]$ being the partition function and $\hat H$ the Hamiltonian of the free field. We recall that the Hamiltonian is time-independent before the motion of the boundaries starts. Note that it is not possible in general for infinite volume systems to write down Gibbs states, as the partition function may not be well-defined since the Hamiltonian is not of trace class. In this case, another approach must be taken to represent the notion of a thermal state, for instance introducing the KMS conditions (see for instance Chapter 22 of~\cite{Strocchi2021}). However, in our case since the spatial direction is bounded because of the mirrors, the Hamiltonian is of trace class and we can use interchangeably thermal and Gibbs states, and the KMS condition is also obeyed.

We can expand the Gibbs density matrix  $\hat \rho_T$ in Fock states of the field at $\tau<\tau_i$, associated with the initial modes given by~\eqref{Eq:initial_modes}, as we have discussed before. The two-point functions of the operators annihilating the {\it in} vacuum characterize this thermal state since 
\begin{equation}
\tr \left[ \hat \rho_{T} \hat a_I^{\dagger} \hat a_J \right] = \frac{\delta_{IJ}}{e^{E_I/T}-1}.
\label{thermal_number}
\end{equation}
Now, for $\tau>\tau_i$ the boundaries start to move in such a way that system will no longer remain at thermal equilibrium at very late times. In order to see this we proceed as before, namely, we compute the number of particles that an asymptotic observer would detect at late times. This is given by the expression
\begin{equation}
     \ev{\hat N_I}_{T}= \tr \left[ \hat \rho_{T} \hat b_I^{\dagger} \hat b_I\right],
\end{equation}
that we can express in terms of the $\alpha$ and $\beta$ coefficients as 
\begin{align}
    \ev{\hat N_I}_{T}=\sum \limits_{J}\abs{\beta_{JI}}^2+\sum \limits_{J}(\abs{\alpha_{JI}}^2+\abs{\beta_{JI}}^2)\frac{1}{e^{ E_J/T}-1}.
\end{align}
This expression has the expected limit as $T\rightarrow 0$ for an initial vacuum state, i.e. it reduces to the one computed in Eq.~\eqref{Eq:number_vacuum}. On the other hand, at high temperatures, the final state will be nearly thermal as long as $\abs{\alpha_{JI}}^2\sim \delta_{IJ}$ and $\abs{\beta_{JI}}^2 \ll 1$. This is what we observe in our numerical simulations (see next section) as long as the amplitudes  $\epsilon_1$ and $\epsilon_2$ are small, and the duration of the bump function \eqref{eq:bump} is short enough. However, we have probed configurations where the mirrors can actually induce a strong mode-mixing and particle production in the final state such that some modes will not follow a thermal distribution, even in the limit of very large temperatures. 

\section{Solving the dynamics: numerical tools}
\label{Sec:Solution}

In this section we will sketch the numerical method that we use to solve the dynamics of the system and compute observables. We will then apply them to some particular configurations of the system that have been considered before in the literature \cite{LI200227,Ling_2002}, showing that our findings are in very good agreement with the results reported there. For the sake of completeness and comparison, we present the method that is used in those references in~ Appendix~\ref{Subsec:Conformal_Transformation_Method}. The method that we adopt is based on finite differences. In the general case, all the modes are coupled among themselves and a truncation to consider a finite number of modes is required to solve the equations numerically. We optimize this approach by performing a Richardson extrapolation to the infinite tower of modes. Both of the approaches lead to results that are in excellent agreement, although the uncertainty of the finite difference method approach combined with the Richardson extrapolation can be kept under control in an easier way. 

The adaptive finite difference methods that we use to solve Eqs.~\eqref{eq:dotphi1d} and~\eqref{eq:dotpi1d} are similar to the ones presented in~\cite{Ruser_2005, Ruser:2006xg, Ruser:2005xg, 2017PhRvE..96a3307V}. We are solving the equations with the initial conditions introduced in Eq.~\eqref{Eq:initial_modes}.  However, there are some differences that are worth remarking. First of all, we adopt an explicit embedded Prince-Dormand (8,9) method of the GNU scientific library, which belongs to the family of the Runge-Kutta methods. These algorithms carry two types of errors (the absolute and the relative ones) whose value can be specified. In our simulations we set the absolute error to be between $10^{-10}$ and $10^{-12}$ and the relative error to zero. Besides, we compute our solutions for $N=128$, $N=256$ and $N=512$ modes, and extrapolate to the limit $N\to\infty$ adopting a Richardson extrapolation. It allowed us to check that our methods have a good convergence in that asymptotic limit. 

Moreover, in order to test the accuracy of each run with a fixed number of modes $N$, we compute the basis of solutions ${\bf u}^{(I)}(\tau)$ with initial data given in Eq. \eqref{Eq:initial_modes}, and compute at all times the inner products in Eq. \eqref{eq:basis} as well as the closure conditions in Eq. \eqref{eq:closure-cond}. To track the errors induced only by the integration of the truncated set of equations, we compute the following error indicators
\begin{widetext}
\begin{eqnarray}\label{eq:Deltas}
    \Delta^{(1)}(\tau) &=& \sqrt{\frac{1}{N^2}\sum_{I,J=1}^N\left(\langle {\bf u}^{(I)}(\tau), {\bf u}^{\left(J\right)}(\tau)\rangle-\delta^{I J}\right)^2}, \\ \nonumber
    \Delta^{(2)}(\tau) &=& \sqrt{\frac{1}{N^2}\sum_{n,m=1}^N\left(\frac{i}{2} \sum_{I=1}^{\infty}\left(-u_{n}^{(I)}(\tau) \bar u_{m}^{(I)}(\tau)+\bar{u}_{n}^{(I)}(\tau) u_{m}^{(I)}(\tau)\right)  - \Omega_{nm}\right)^2}.
\end{eqnarray}
\end{widetext}
Below we will provide precise numbers for them. 

Let us now consider some concrete examples. In particular, we will assume that the boundaries follow the simple damped oscillatory trajectories 
\begin{eqnarray}\label{eq:fg-traj}
    g(t) &=& 1 + 
   \epsilon_1 B(t) (\sin(q\pi t + \phi) - 
      \sin(\phi)), \\ \nonumber
f(t) &=&  \epsilon_2 B(t)
   \sin(q \pi t)  , 
\end{eqnarray}
in coordinates $(t,x)$, where $\epsilon_1$ and $\epsilon_2$ control the amplitudes of the oscillations of the boundaries, $q$ is an integer that controls their frequency, $\phi$ is a relative phase, and $B(t)$ is the bump function  
\begin{align}
    B(t)= \begin{cases}
      e^{\frac{1}{\sigma}\left(1-\frac{1}{(1 - (t/\Gamma - 1)^2)}\right)}  & t \in (0, 2 \Gamma) \\
      0 & t \notin (0, 2 \Gamma)
    \end{cases}, 
\label{eq:bump}
\end{align}
with the parameters $\sigma$ and $\Gamma$ controlling the steepness and duration of the bump function. This is a suitable function to represent the motion in a finite interval of time $t \in (0, 2 \Gamma)$ since the function identically vanishes for $t \notin (0, 2 \Gamma)$ and it is $C^{\infty}$ everywhere. For simplicity, we report here only the examples in which $ \Gamma = 1$, and hence the function $B(t)$ will be identically zero for $t \leq 0$ and $t \geq 2$. Thus, the dynamics outside such interval is trivial. The other choices that we have studied for which the trajectory looks qualitatively the same have lead also to similar results.
\subsection{One moving boundary: $\epsilon_{2} = 0$.} \label{subsec:one-mov}
Let us start with the left boundary at rest by setting $\epsilon_2=0$, namely, $f(t) = 0$. Besides, let us consider that the other boundary oscillates with frequency given by $q = 10$, amplitude $\epsilon_1 = 1/40$ and phase $\phi=\pi$. Moreover, in the bump function we set $\sigma=1/10$, namely, the boundary  oscillates with nearly constant amplitude in the interval $t\in[0,2]$ several times, and remains at rest at $g(t) = 1$ for $t\leq 0$ and $t \geq 2$. We set the field in the natural static vacuum state at times $t\leq 0$. After the boundary becomes at rest again at $t\geq 2$, the system is not in the natural vacuum state at late times.
\begin{figure}[ht]
{\centering     
  \includegraphics[width = 0.49\textwidth]{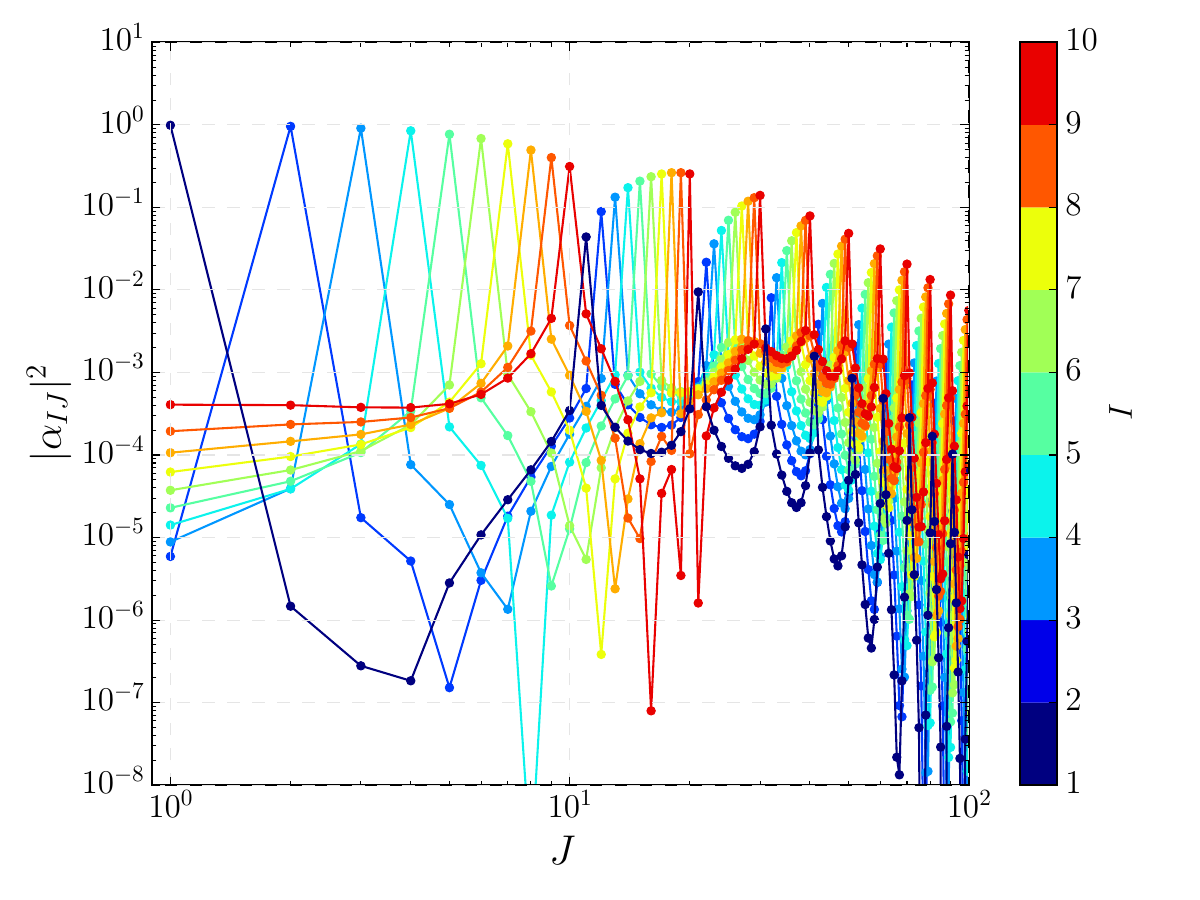}
  \includegraphics[width = 0.49\textwidth]{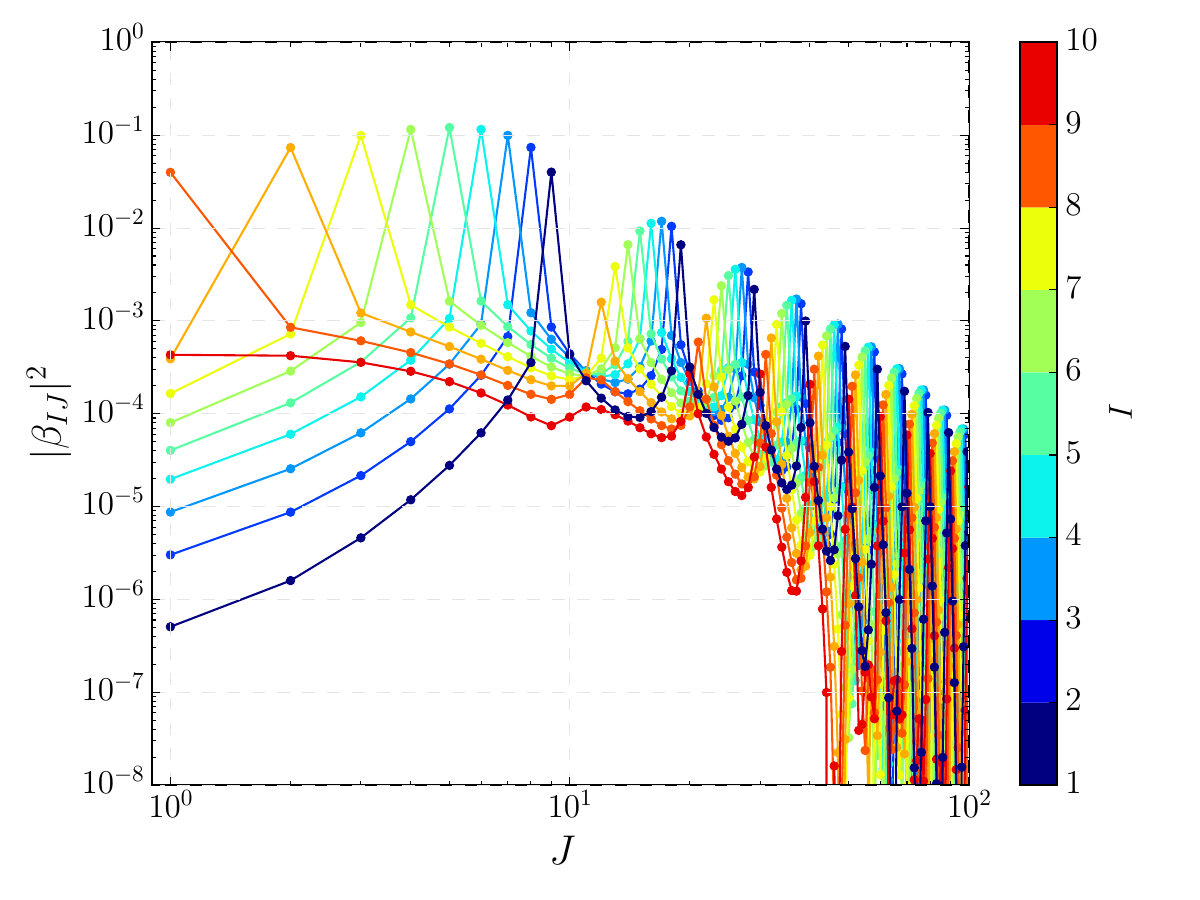}
}
\caption{Bogoliubov coefficients: Upper panel: We show the modulus squared of the $\alpha_{IJ}$ coefficient. Lower panel: We plot the modulus squared of the $\beta_{IJ}$ coefficient. These plots correspond to the trajectories given by Eqs. \eqref{eq:fg-traj} and \eqref{eq:bump}, with $\epsilon_1 = 1/40$, $\epsilon_2 = 0$, $q = 10$, $\phi=\pi$ and $\sigma=1/10$.}
\label{fig:beta1}
\end{figure}
Actually, in Fig. \ref{fig:beta1} we show the modulus of some of the Bogoliubov coefficients relating the $in$ and $out$ states. We clearly see that the coefficients reach a maximum value at concrete frequencies where resonances occur. For instance, we see that for the solution ${\bf u}^{(I)}\left(\tau\right)$, with $I=1$, which amounts the solution exciting the mode $n=1$ in the past, at late times, the Bogoliubov coefficients $\alpha_{IJ}$ reach local maxima (resonances) at frequencies $I(p)=p_1q+J$, with $p_1$ being a natural number and $q$ the integer controlling the frequency of oscillation of the boundary. In these cases, the mode mixing becomes very efficient (this can be seen as a beam splitter transformation). One can see that this pattern is present for $I=2,3,\ldots$. For the $\beta_{IJ}$, we see that the maxima appear instead at $I(p)=p_2q-J$. This implies that particles are created more efficiently in modes satisfying that relation. This resonance structure has already been discussed, for instance, in \cite{Crocce:2001zz,Sabín_2014}. 

\subsection{Two moving boundaries}\label{sec:2-mov-boundar}
We have also studied the case in which the two boundaries move. Here, we set $\epsilon_1=\epsilon_2=1/40$. We set the frequencies of the boundaries  $q = 10$. For the bump function we set $\sigma=1/10$. We start with the case $\phi=\pi$, namely, the boundaries oscillate with opposite phases. One can see that in this case $\dot L=2\dot f$. 
\begin{figure}[ht]
{\centering     
  \includegraphics[width = 0.49\textwidth]{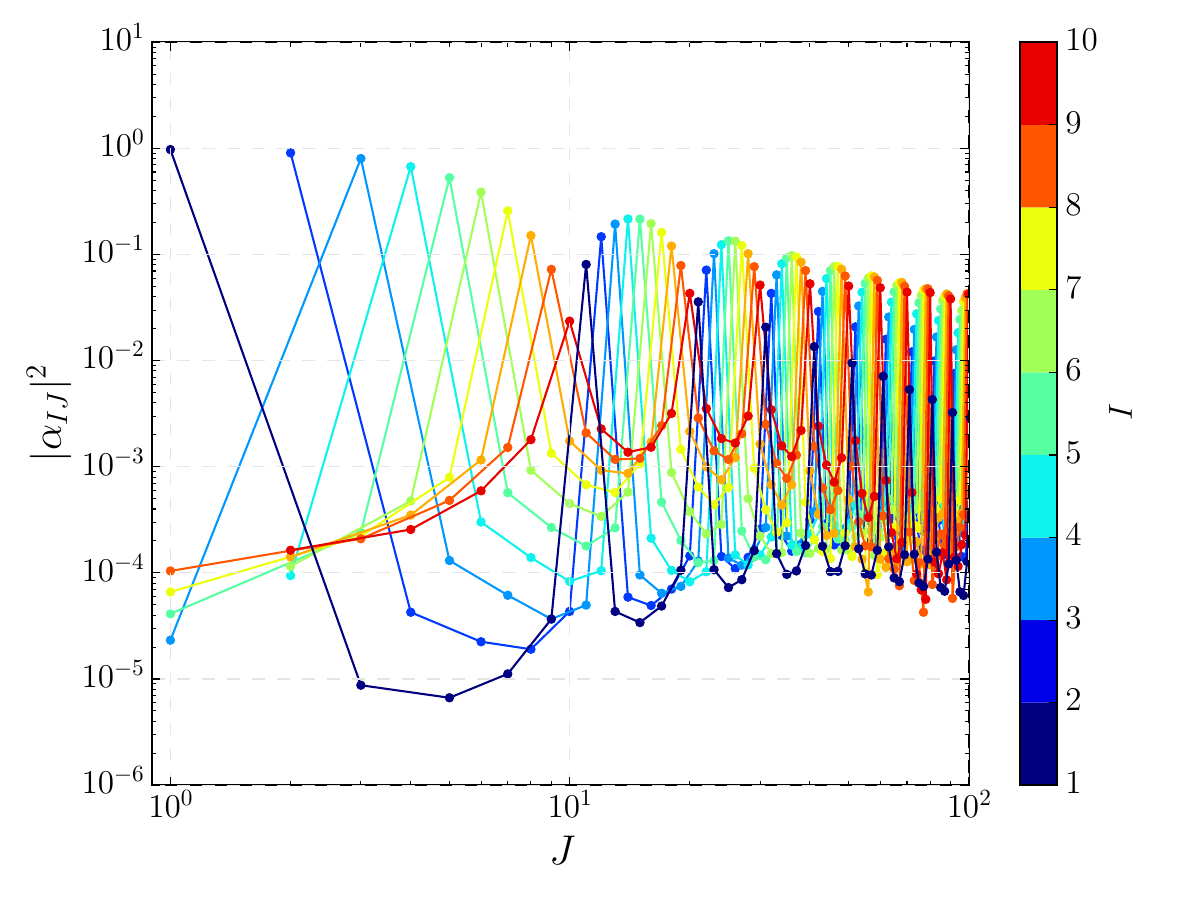}
  \includegraphics[width = 0.49\textwidth]{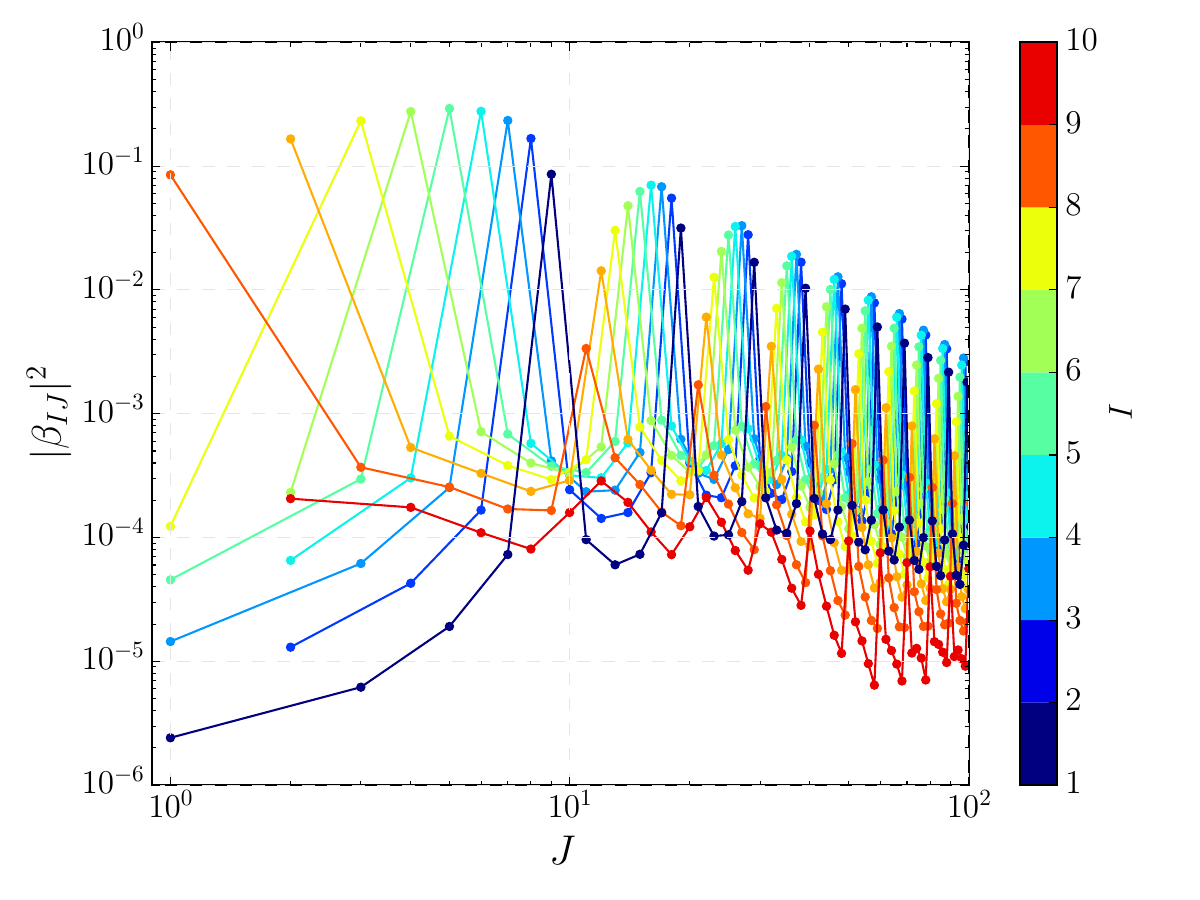}
}
\caption{Bogoliubov coefficients: Upper panel: We plot the modulus squared of the $\alpha_{IJ}$ coefficient. Lower panel: We show the modulus squared of the $\beta_{IJ}$ coefficient. These plots correspond to the trajectories given by Eqs. \eqref{eq:fg-traj} and \eqref{eq:bump}, with $\epsilon_1 =\epsilon_2= 1/40$, $q = 10$, $\phi=\pi$ and $\sigma=1/10$.}
\label{fig:beta2}
\end{figure}
In Fig. \ref{fig:beta2} we see again a resonance structure similar to the previous case, with only one boundary moving. However, in this case, it is interesting to note that $|\alpha_{IJ}|^2$ and $|\beta_{IJ}|^2$ are negligible whenever $I+J$ is odd. This is a new feature with respect to the previous case. This is direct consequence of equations of motion \eqref{eq:dotphi1d} and \eqref{eq:dotpi1d}. Concretely, by direct inspection, one can show that the even and the odd modes decouple as $\dot{L} \to  2\dot{f}$. Besides, we have seen that in general, whenever $q$ is an even number, we see a similar structure regarding mode mixing and particle production. Nevertheless, if $q$ is odd, mode mixing and particle production turns out to be less efficient. 

We have also studied the case in which the two boundaries oscillate in phase, namely, $\phi=0$. In this case, $\dot L=0$. The result is depicted in Fig.~\ref{fig:beta3}. We appreciate also a resonance structure in both the $|\alpha_{IJ}|^2$ and the $|\beta_{IJ}|^2$ coefficients. In this case, the particle production is quantitatively smaller than in the two previous cases. One can see, for instance, that the diagonal coefficients $|\alpha_{II}|^2\simeq 1$, while the off-diagonal coefficients are bounded by $|\alpha_{IJ}|^2 \lesssim 1/10$, when $I\neq J$, decaying fast to zero as $\abs{I-J}$ increases. We see this behavior for several choices of $q$ when it is an even number. However, if $q$ is odd, there is a qualitatively stronger mode mixing and particle production.
\begin{figure}[ht]
{\centering     
  \includegraphics[width = 0.49\textwidth]{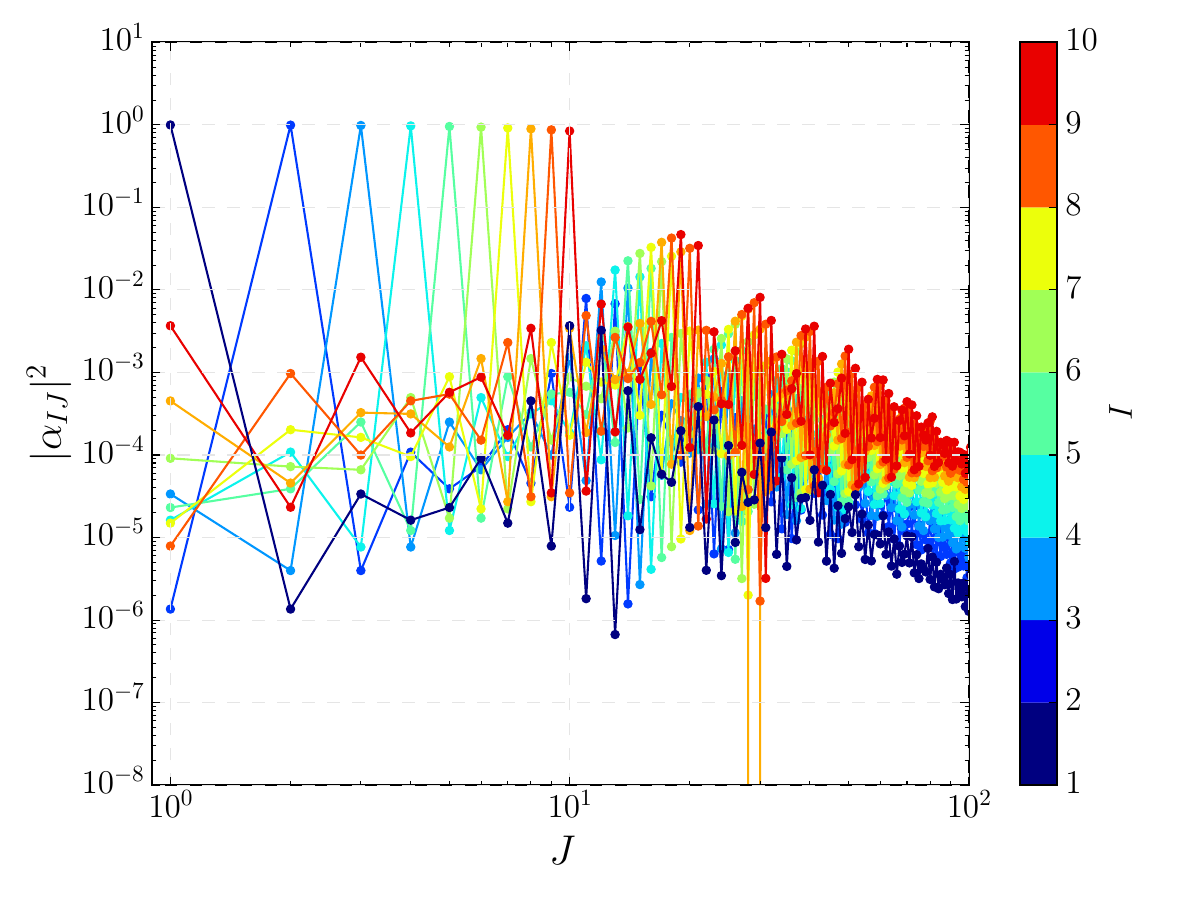}
  \includegraphics[width = 0.49\textwidth]{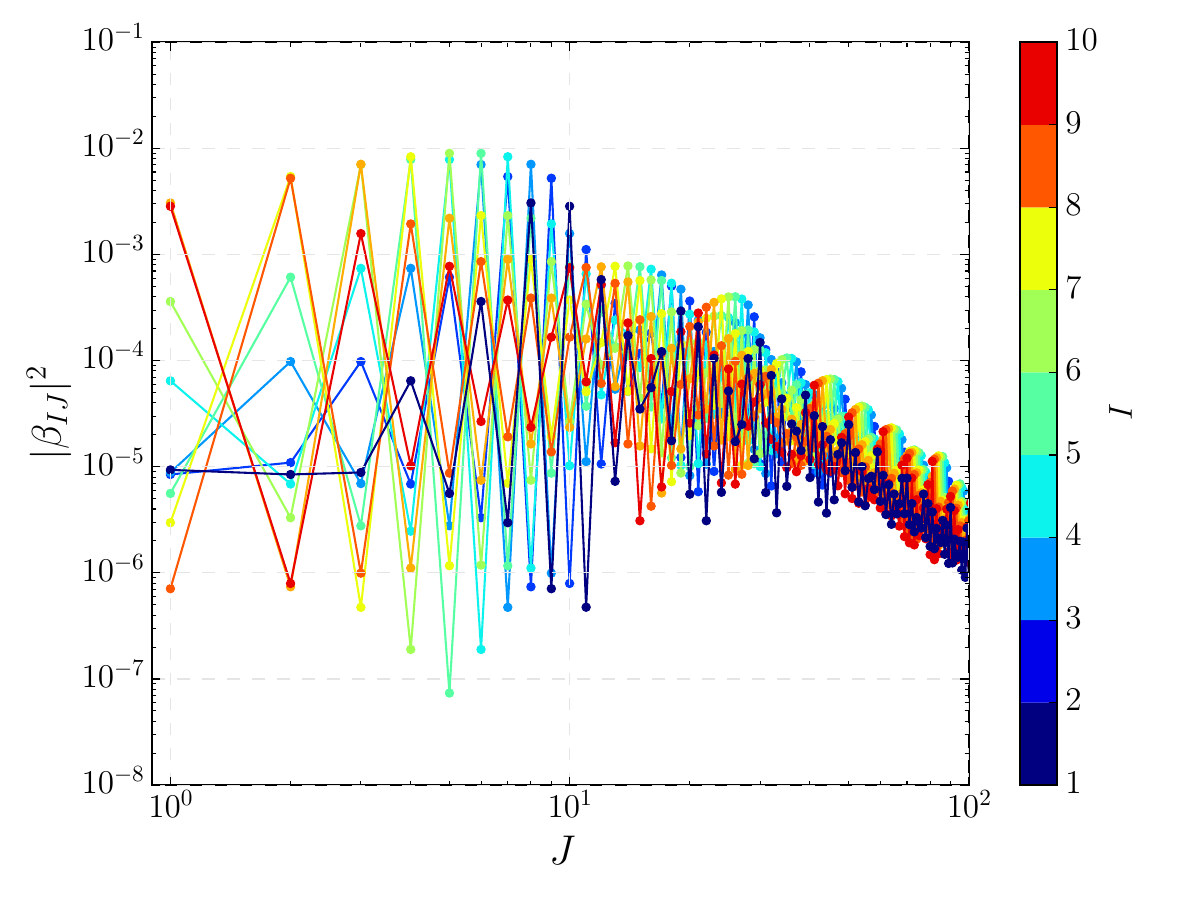}
}
\caption{Bogoliubov coefficients: Upper panel: We show the modulus squared of the $\alpha_{IJ}$ coefficient. Lower panel: We plot the modulus squared of the $\beta_{IJ}$ coefficient. These plots correspond to the trajectories given by Eqs. \eqref{eq:fg-traj} and \eqref{eq:bump}, with $\epsilon_1 = \epsilon_2 = 1/40$, $q = 10$, $\phi=0$ and $\sigma=1/10$.}
\label{fig:beta3}
\end{figure}

\subsection{Particle production for a thermal state}

We have also studied the effects of mode-mixing and particle production on initially populated states following a thermal distribution with temperature $T$, as defined in Eq. \eqref{eq:thermal-state}. For the trajectories already considered, we have computed the particle number $\langle\hat N_I\rangle_{T}$ once the mirrors become at rest at late times. 
In Fig.~\ref{fig:thermal} we show this observable for two different configurations of the mirrors and several initial temperatures. 
\begin{figure}[ht]
{\centering
  \includegraphics[width = 0.49\textwidth]{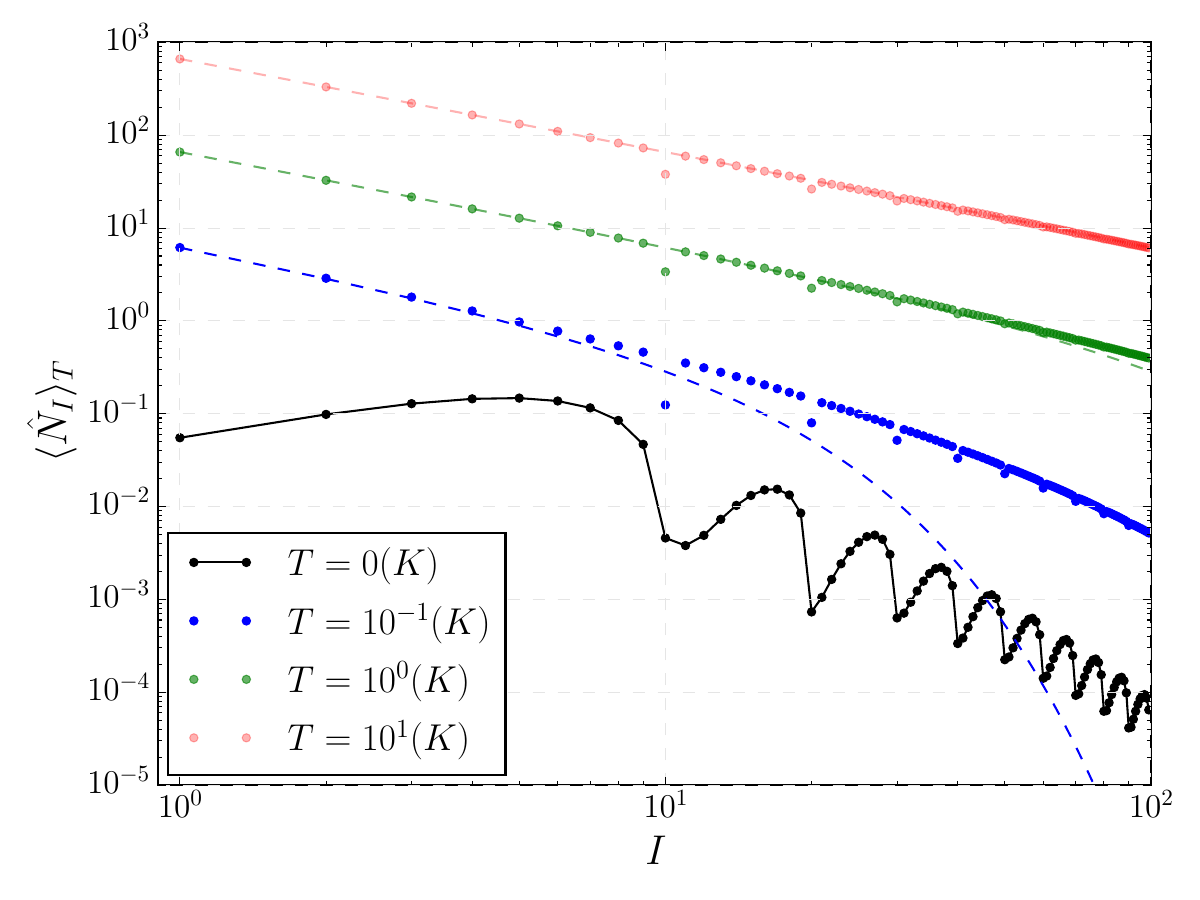}
  \includegraphics[width = 0.49\textwidth]{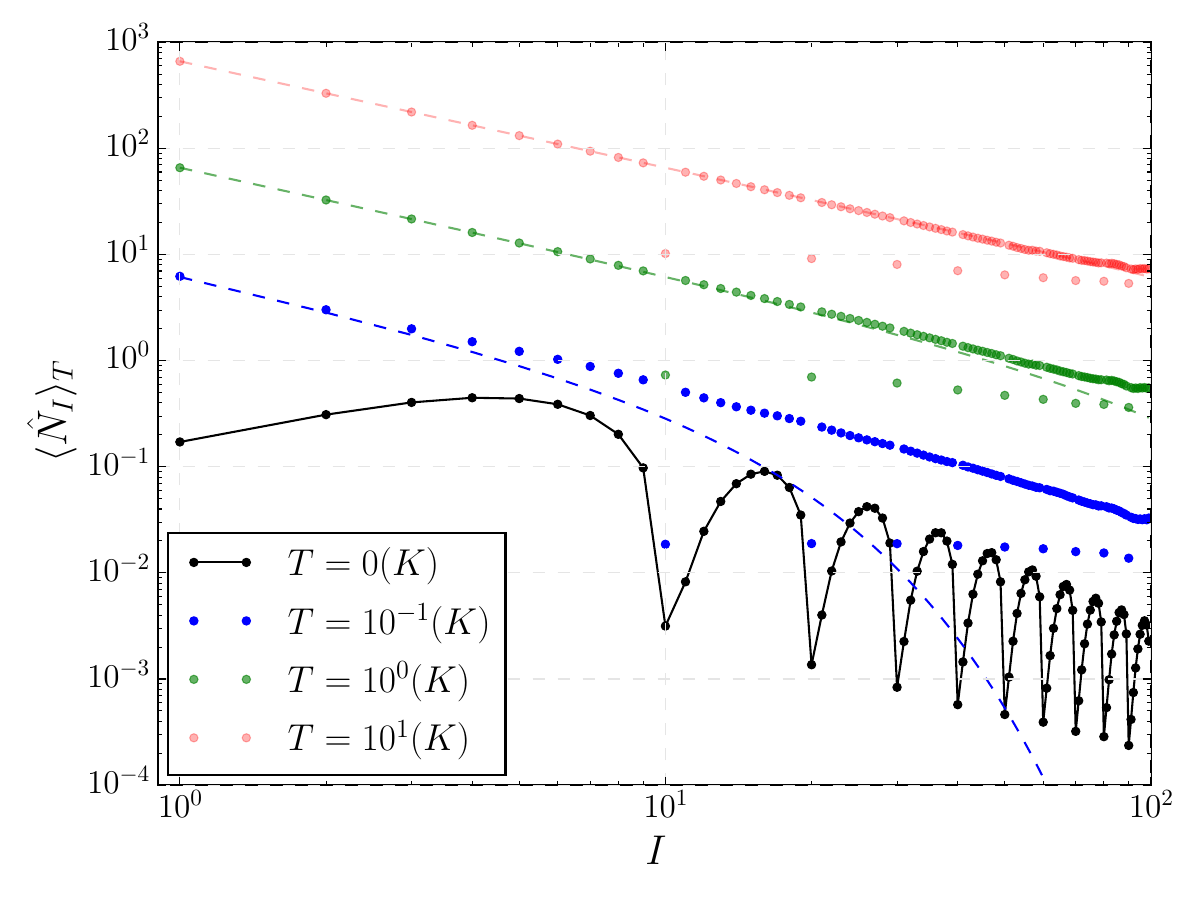}
}
\caption{Particle number: We show the particle number $\langle\hat N_I\rangle_{T}$ for an initial thermal state with different temperatures, and the simulation corresponding to the trajectories given by Eqs. \eqref{eq:fg-traj} and \eqref{eq:bump}, with $q = 10$, and $\sigma=1/10$. Upper panel: Here we set $\epsilon_1 = 1/40$, $\epsilon_2 = 0$, and $\phi=\pi$. Lower panel: Here we set $\epsilon_1 = \epsilon_2 = 1/40$, and $\phi=\pi$.}
\label{fig:thermal}
\end{figure}
In these cases, despite most of the infrared modes remain almost in a thermal state, there are (anti)resonant modes with less occupation number, regardless of the value of the initial temperature. This is to be compared with what occurs with high frequency modes: for those modes we see that their population is higher than in a thermal state, but the spectrum still exhibits some (anti)resonant structure for some modes that are less populated than what a naive interpolation taking their neighbors as data would predict. This phenomenon can be explained by looking at the non-obvious behavior of the Bogoliubov coefficients. In general, we observe that the mode-mixing and the particle creation of the (anti)resonant modes, i.e. $I=r\,q$ with $r$ a positive integer, conspire in such a way that the particle number in those modes $I$ do not follow a nearly thermal distribution. For instance, let us consider the resonant mode $I=10$. In Fig.~\ref{fig:beta4}, we see that for $I\neq 10$ and $I\sim 10$, and $J\leq I$, either $|\alpha_{IJ}|^2$ or $|\beta_{IJ}|^2$ are always one order of magnitude larger than those coefficients for $I=10$, the (anti)resonant mode. We have seen that this behavior is general for all the other (anti)resonant modes $I = 20,30,\ldots$ This particular behavior of the Bogoliubov coefficients of the (anti)resonant modes is the responsible for the particle number to be below the naive expectation. 
\begin{figure}[ht]
{\centering
  \includegraphics[width = 0.49\textwidth]{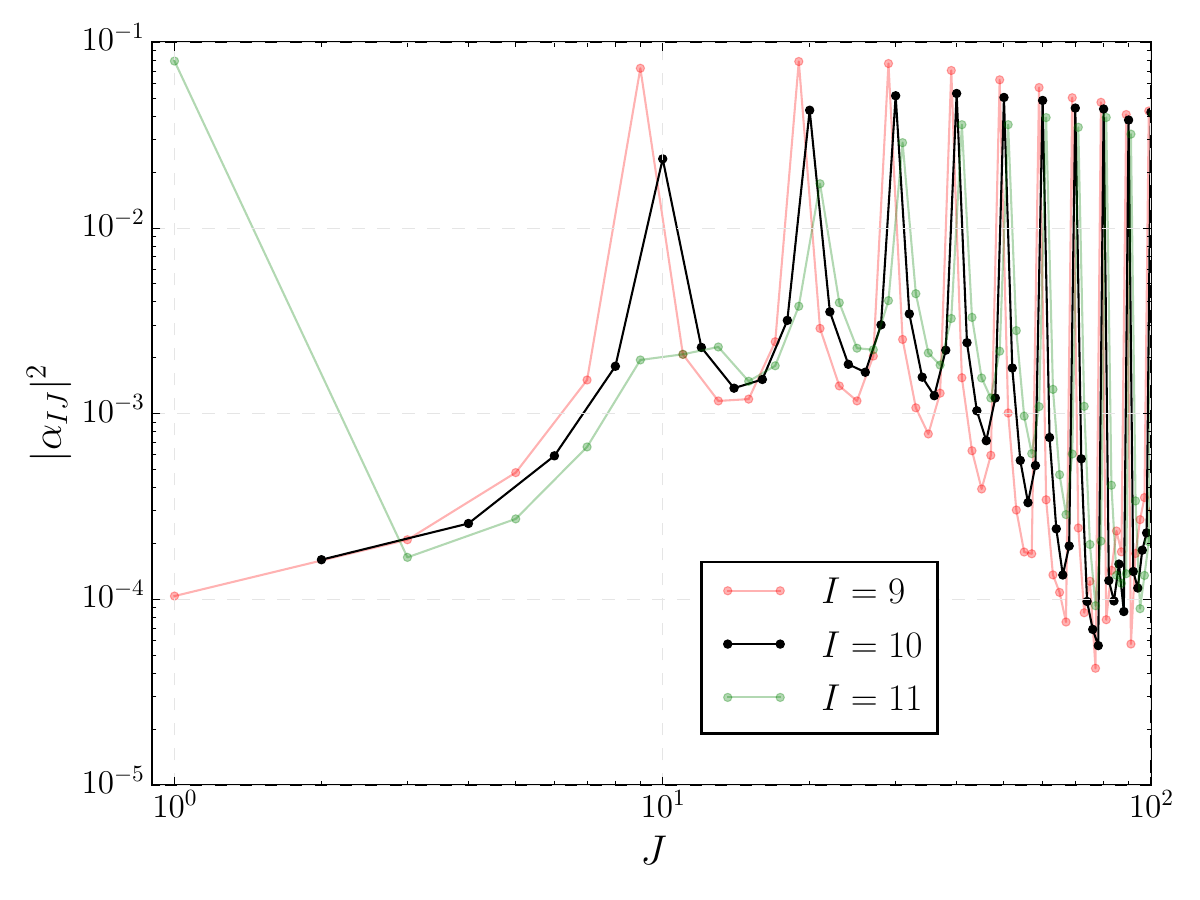}
  \includegraphics[width = 0.49\textwidth]{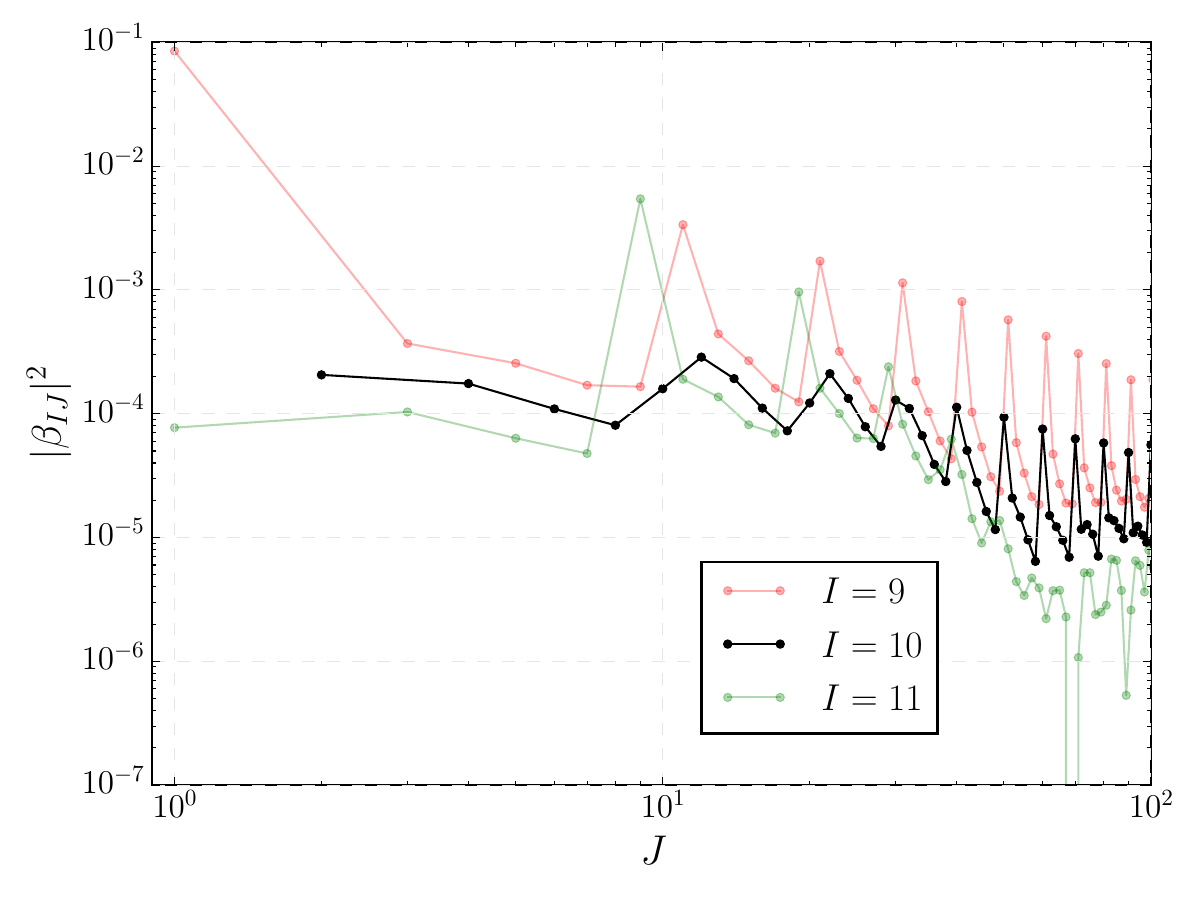}
}
\caption{Bogoliubov coefficients: We show the Bogoliubov coefficients for the modes $I=9,10,11$ for the simulation corresponding to the trajectories given by Eqs. \eqref{eq:fg-traj} and \eqref{eq:bump}, with $q = 10$, $\sigma=1/10$, $\epsilon_1 = \epsilon_2 = 1/40$, and $\phi=\pi$. Upper panel: $\alpha$-coefficient. Lower panel: $\beta$-coefficient.}
\label{fig:beta4}
\end{figure}
In order to confirm that this is the case, we have carried out simulations where the cavities oscillate away from resonant frequencies, i.e. $q$ not being an integer. Actually, the less resonant cases are those in which $q$ is a semi-integer. Concretely, we have analyzed the cases $q=2.5$, $q=4.5$ and $q=10.5$. Although in these cases we still observe a strong mode mixing, it does not manifest in the occupation number of modes close to the frequency of oscillation since they remain close to their thermal state value. The UV modes behave in a similar way: their occupation number increases considerably (they ``warm up''). Hence, in the case of non resonant frequencies of the cavities, the mode mixing is not as efficient to cool down the infrared modes as in the resonant cases.

We also want to note that the spontaneous creation of particles ($T=0$) remains several orders of magnitude below the occupation number of the (anti)resonant modes, and therefore its contribution to the total particle production is negligible. As a final comment, the configuration of the mirrors with $\epsilon_1 = \epsilon_2 = 1/40$, $q = 10$, $\phi=0$ and $\sigma=1/10$ does not show such a strong mode-mixing, and the Bogoliubov coefficients behave nearly as $\abs{\alpha_{JI}}^2\sim \delta_{IJ}$ and $\abs{\beta_{JI}}^2 \ll 1$ (see Fig.~\ref{fig:beta3}). This is not the case for this configuration of the boundaries if $q$ is an odd number. Similarly, for the configuration $\epsilon_1 = \epsilon_2 = 1/40$, $\phi=\pi$, $\sigma=1/10$ and $q$ any odd number, the mode mixing is weak and the Bogoliubov coefficients behave nearly as $\abs{\alpha_{JI}}^2\sim \delta_{IJ}$ and $\abs{\beta_{JI}}^2 \ll 1$, as shown in Fig.~\ref{fig:beta3}.

\subsection{Error estimation}

As a final remark, we analyze the accuracy of our numerical simulations. The strategy we adopt here is the following. First, we compute the basis of complex solutions ${\bf u}^{(I)}\left(\tau\right)$ at all times $\tau$ for the truncated theory, i.e. we consider a finite number of modes $N$ that is sufficiently large. Then, we compute the error indicators introduced in Eq. \eqref{eq:Deltas} at all times. 
In Fig.~\ref{fig:deltas} we show the results for the simulation carried out in Subsec.~\ref{subsec:one-mov}, namely, the case in which one of the boundaries is at rest in coordinates $(t,x)$. The errors we obtain when computing these quantities for simulations involving $N=256$ modes are always well below $10^{-9}$ at all times. This can be seen in Fig. \ref{fig:deltas}.
\begin{figure}[ht]
{\centering     
  \includegraphics[width = 0.49\textwidth]{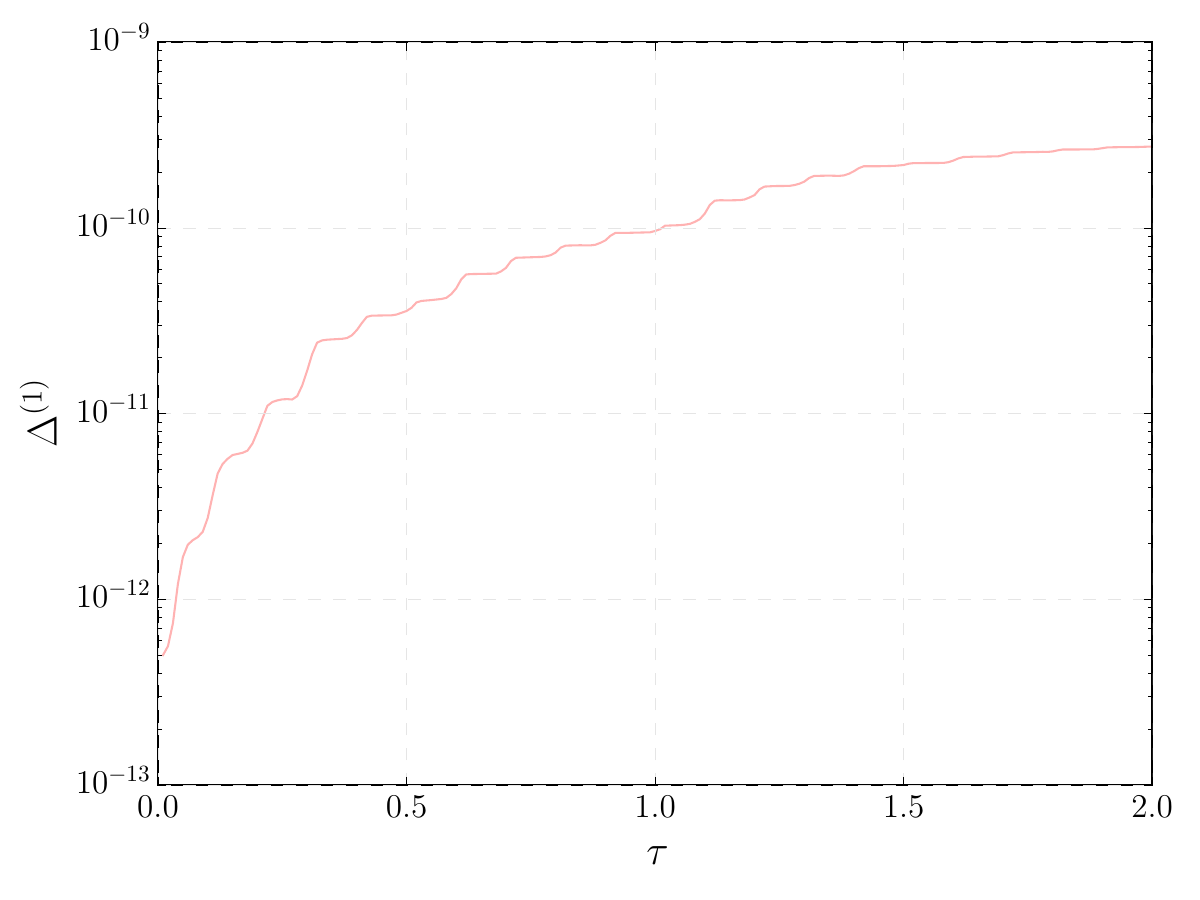}
  \includegraphics[width = 0.49\textwidth]{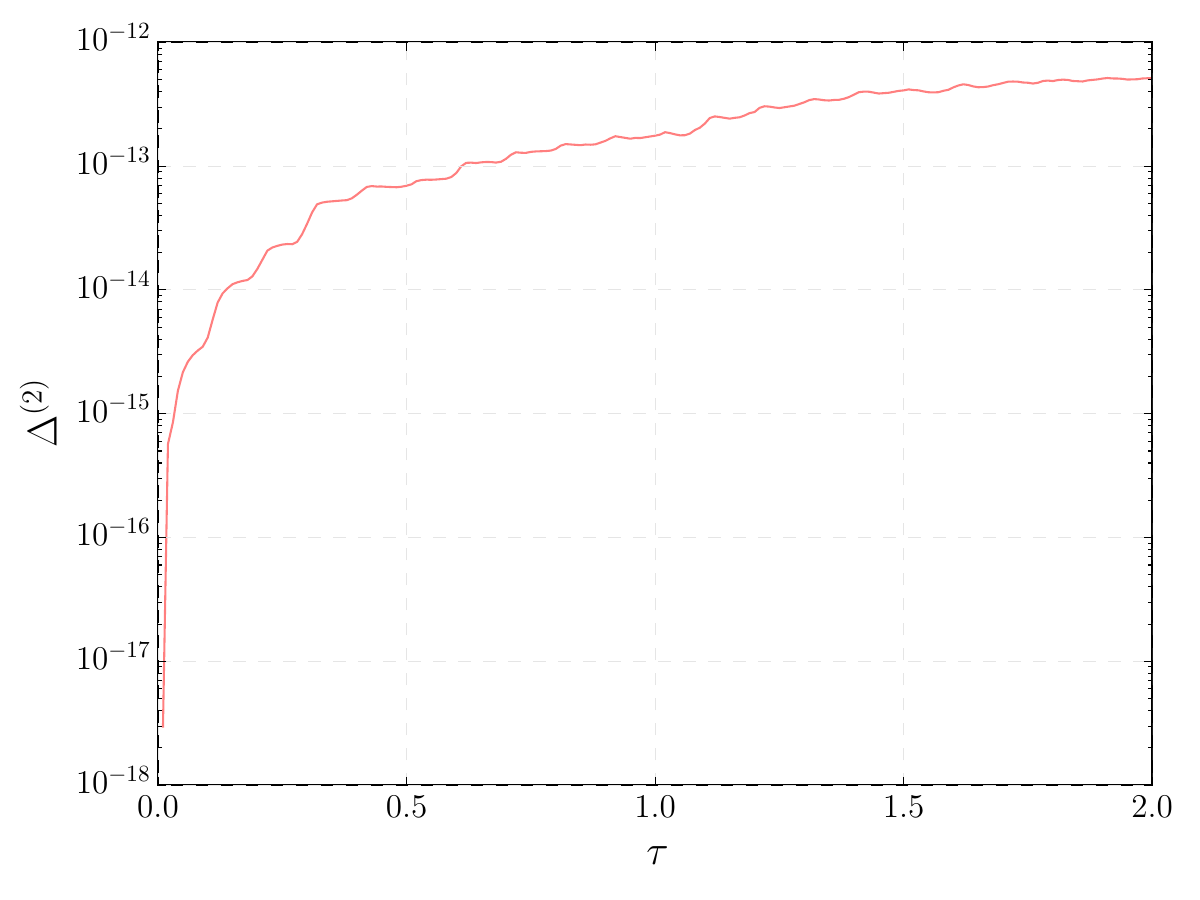}
}
\caption{Error indicators in Eq. \eqref{eq:Deltas} for the simulation corresponding to $N=256$ and for the configuration of the mirrors studied in Subsec. \ref{subsec:one-mov}. Upper panel: Numerical values of the indicator $\Delta^{(1)}$. Lower panel: Numerical values of the indicator $\Delta^{(2)}$. The error does not increase for $\tau >2$ since all the non-trivial dynamics occurs for $\tau \in (0,2)$.}
\label{fig:deltas}
\end{figure}

\section{Conclusions}
\label{Sec:conclusions}
In summary, we have shown that a (1+1)-dimensional field theory in Minkowski spacetime confined in a cavity with moving boundaries is in one to one relation with a (1+1)-dimensional field theory on a (flat) acoustic metric with static boundaries. To the best of our knowledge, this analogy has not been suggested before. Besides, we provide a detailed quantum theory of the model, where the dynamics is solved numerically in nonperturbative regimes, e.g. situations where the speed and accelerations of the boundaries are relativistic. Our numerical tools allow to compute all the relevant observables at late times, providing precise and accurate results. Among the most interesting ones, we have found that, in the nonperturbative regime and for initially populated thermal states, there are configurations of the mirrors producing a strong mode-mixing and particle production that makes some (anti)resonant modes to lose particles in their occupation number, reaching a final state that is not thermal, regardless of the  temperature of the initial state. This phenomenon has already been reported in the literature, for instance, in Refs. \cite{DODONOV1996219,Dodonov:2020eto}, as an effective cooling down of the infrared modes. Strictly speaking, denoting this decreasing in the occupation numbers of some modes as an ``effective cooling down'' is not accurate, since the final state is not thermal and has no temperature associated with. However, we use this nomenclature for simplicity. Besides, it has only been discussed for concrete configurations of the mirrors: i) one of the mirrors remaining static at all times, and ii) for  particular resonant frequencies of oscillation of the cavity (corresponding to its fundamental mode and the first overtone). Here, we show that the effective cooling down reported in \cite{Dodonov:2020eto} appears in higher resonant frequencies, and also when the two mirrors oscillate in phase opposition. 

The content of this paper provides the theoretical grounds and the numerical methods that we have used to obtain the results achieved in \cite{GarciaMartin-Caro:2023jjq}. Here, we have applied our tools to the case of oscillating boundaries, either only one or both of them.

A comment regarding the relevance of the analysis presented here for experimental proposals is in order. In Ref. \cite{Johansson2010} it is proposed an experimental setting that probes the dynamical Casimir effect induced on an electric field propagating in a coplanar wave guide (CPW) by means of effective boundaries due to superconducting interference devices (SQUIDs) that can oscillate with frequencies smaller than $\omega_p = 37.3$~GHz, which corresponds to the plasma frequency of the SQUIDs. The phase field inside these cavities fulfills Robin boundary conditions. However, in the approximation in which $\delta L \ll L$, or equivalently, for modes satisfying $\lambda \gg \delta L$, Robin boundary conditions are well approximated by Dirichlet boundary conditions, the ones we consider in our manuscript. Hence, our results straightforwardly apply whenever these approximations are valid. Moreover, let us also note that the length of the CPW in these experiments can be as large as $L_0=10.0$~cm and the change in size can be as large as $\delta L=0.25$~cm. In addition, the speed of propagation of the phase field (the time integral of the electric field) propagating in the CPW is $v\simeq 10^{10}$~(cm/s). With all this in mind, one can easily see that the (driving) frequency of the oscillation of the boundaries in our simulations will correspond to $\omega_d = 31.4$~GHz. This clearly implies $\omega_d \sim \omega_p$. Therefore, the configurations explored in this manuscript are close but exceed the experimental capabilities, that require $\omega_d < \omega_p$. 
One way out of this limitation is to work with waveguides with smaller speeds of propagation $v$ of the phase field, by one or two orders of magnitude, where one could work with driving frequencies fulfilling $\omega_d<\omega_p$. However we stress that our numerical tools can still be used to analyze the experimental setups proposed in the literature, e.g. the one in Ref.~\cite{Johansson2010}.

Finally, regarding future directions of work, we are currently working on extending our methods to other boundary conditions, namely Neumann boundary conditions or even the most general Robin boundary conditions. They are specially relevant for the waveguides systems that we have discussed above. It will be also interesting to explore geometries of the boundaries that are different from the ones presented in the Appendix~\ref{Subsec:3+1} (which are the ones for a parallelepiped) for example, those of a cylinder (which are relevant for some waveguides) or a sphere. Another direction that is worth exploring is the study of the entanglement generated in the process (in the lines of Ref.~\cite{deOliveira2023}, specially for the resonant trajectories. In fact, it would be particularly interesting to study the resilience of the entanglement generated against thermal noise in the system, as well as other aspects concerning the thermodynamical properties of the quantum vacuum \cite{Xie1}. Finally, we plan to investigate possible cooling protocols for the infrared sector of the system, something that could be useful to enhance some of the quantum aspects of the phenomenon.

\acknowledgments

The authors would like to thank Carlos Barcel\'o and Ignacio Reyes for helpful discussions. Financial support is provided by the Spanish Government through the projects PID2020-118159GB-C43, PID2020-119632GB-I00, and PID2019-105943GB-I00 (with FEDER contribution). AGMC acknowledges financial support  from the PID2021-123703NB-C21 grant funded by MCIN/ AEI/10.13039/501100011033/ and by ERDF, ``A way of making Europe''; and the Basque Government grant (IT-1628-22). GGM is funded by the Spanish Government fellowship FPU20/01684 and acknowledges financial support from the grant CEX2021-001131-S funded by MCIN/AEI/10.13039/501100011033. JO is supported by the ``Operative Program FEDER2014-2020 Junta de Andaluc\'ia-Consejer\'ia de Econom\'ia y Conocimiento'' under project E-FQM-262-UGR18 by Universidad de Granada. JMSV acknowledges the support of the Spanish Agencia Estatal de Investigaci\'on through the grant “IFT Centro de Excelencia Severo Ochoa CEX2020-001007-S".

\appendix

\section{Conformal transformation method}
\label{Subsec:Conformal_Transformation_Method}
Let us introduce the so-called conformal transformation method in which we reduce the problem to solving Moore's equations~\cite{Moore1970}. This can be reduced to an algebraic problem of finding the roots of some equations, and hence it is possible to solve the dynamics in a simple way. 

The approach that we will describe here to solve the system
\begin{align}
    & \left( \partial_t^2 - \partial_x^2 \right) \phi(t,x) = 0, \nonumber \\
    & \phi(t, f(t))  =  \phi(t,g(t)) =0,
    \label{Eq:Klein-Gordon}
\end{align}
is based on a conformal transformation to a new set of coordinates $(s,w)$ :
\begin{equation}
    s + w = G(t+x),\qquad s-w = F(t-x)
\end{equation}
which preserves the form of the equation of motion for the scalar field~\eqref{Eq:Klein-Gordon} up to a conformal factor,
\begin{equation}
    (\partial_t^2-\partial_x^2)\phi=F'G'(\partial_s^2-\partial_w^2)\phi=0,
\end{equation}
with $F',G' \neq 0$, i.e., both $F$ and $G$ need to be monotonic functions on their argument. We choose the functions $G$ and $F$ so that, in the new coordinates, the Dirichlet boundary conditions become trivial, i.e., independent of the new time coordinate:
\begin{equation}
    \phi(t,f(t))=0=\phi(s,w=0),\;\phi(t,g(t))=0=\phi(s,w=1).
\end{equation}
Such conditions will be satisfied if and only if:
\begin{align}
    w(t,f(t))&\equiv\tfrac{1}{2}[G(t+f(t))-F(t-f(t))]=0,
    \label{Moores1}\\ w(t,g(t))&\equiv \tfrac{1}{2}[G(t+g(t))-F(t-g(t))]=1,
    \label{Moores2}
\end{align}
for all $t$. The previous equations, also known as (generalized) Moore's equations \cite{Moore1970}, allow for the explicit determination of the functions $G$ and $F$ once the trajectories of the mirrors are known. Also, for time-independent Dirichlet boundary conditions, the Klein-Gordon equation is easily solved using the new coordinates. Hence, a complete basis set of solutions for the problem will be given by ${\bf u}^{I}(x,t)=\delta^I_n\,\psi_n(t,x)$, where
\begin{equation}
    \psi_n(t,x)=\frac{i}{\sqrt{4\pi n }}[e^{-i\pi nG(t+x)}-e^{-i\pi n F(t-x)}].
\label{eq:evomodes}
\end{equation}
Indeed, such modes satisfy both the field equation and the boundary conditions.

Notice that although the coordinates $(\tau, \xi)$ introduced in Sec.~\ref{Sec:classical} also trivialize the boundary conditions, they do not preserve the structure of the equations of motion up to a conformal factor. In that sense, the change of coordinates done here is different. Also, finding the change of coordinates in this case is tantamount to solving the classical dynamics, whereas in the $(\tau,\xi)$ the dynamical equations of motion are still highly convoluted. 

The geometrical derivation of $G(z)$ and $F(z)$ is based on tracing back along a sequence of null lines from each point $z=x+t$ or $z=x-t$ until a null line intersects the time axis in the static region $[-\infty,\Lambda_f]$ (for $G$) or $[-\infty,0]$ (for $F$) where these functions can be evaluated directly (see Refs. \cite{PhysRevA.52.4405,PhysRevLett.76.408,LI200227,Ling_2002}).
\begin{figure}
    \centering
    \tikzset{every picture/.style={line width=0.75pt}} 

\begin{tikzpicture}[x=0.75pt,y=0.75pt,yscale=-1,xscale=1]

\draw    (92.93,50.08) -- (93.61,12.97) ;
\draw    (92.93,50.08) .. controls (92.51,73.4) and (82.14,74.86) .. (94,91.98) .. controls (105.86,109.1) and (81.46,112.5) .. (93.03,128.55) .. controls (104.61,144.61) and (96.08,146.02) .. (95.21,177.29) ;
\draw    (95.21,177.29) -- (94.65,225) ;

\draw    (188,172) -- (188.82,222.93) ;
\draw    (188,172) .. controls (188,143) and (198.44,128.44) .. (186,105) .. controls (173.56,81.56) and (184,89) .. (184.16,44.51) ;
\draw    (184.16,44.51) -- (184,15) ;
\draw  [dash pattern={on 0.84pt off 2.51pt}]  (95,185) -- (189.09,185.72) ;
\draw  [dash pattern={on 0.84pt off 2.51pt}]  (93.32,44.34) -- (184.16,44.51) ;
\draw [color={rgb, 255:red, 208; green, 2; blue, 27 }  ,draw opacity=1 ]   (105.79,44.34) -- (180,87) -- (100,142) -- (188.41,197.46) ;

\draw (61.58,10.31) node [anchor=north west][inner sep=0.75pt]    {$f( t)$};
\draw (189.31,8.01) node [anchor=north west][inner sep=0.75pt]    {$g( t)$};
\draw (188.83,35.99) node [anchor=north west][inner sep=0.75pt]    {$t=t_{0}$};
\draw (97.25,22.71) node [anchor=north west][inner sep=0.75pt]    {$( t_{0} ,z-t_{0})$};
\draw (182.84,77.65) node [anchor=north west][inner sep=0.75pt]    {$( t_{1} ,g( t_{1}))$};
\draw (28.84,132.65) node [anchor=north west][inner sep=0.75pt]    {$( t_{2} ,f( t_{2}))$};
\draw (190,175.4) node [anchor=north west][inner sep=0.75pt]    {$( 0,\Lambda )$};
\draw (55.84,175.65) node [anchor=north west][inner sep=0.75pt]    {$( 0,0)$};

\end{tikzpicture}
    \caption{Null ray propagation inside a cavity with oscillating walls.}
    \label{fig:DCEconformal}
\end{figure}
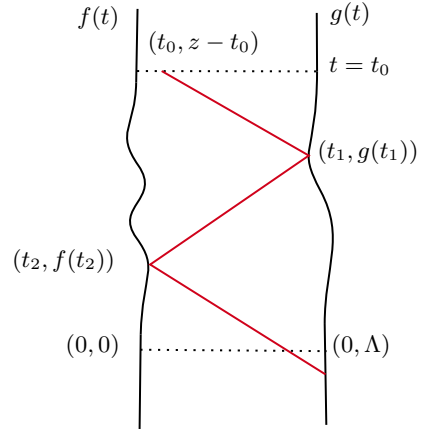
For example, as we illustrate in Fig.~\ref{fig:DCEconformal}, to determine $G(z=x+t)$ at a given point $(x_0,t_0)$, we trace back a ray until it intersects the right mirror at the point $(t_1, g(t_1))$. The value of $t_1$ can be obtained since the light cone coordinate $u=x+t$ is constant along the path of left-moving light rays, so that
\begin{equation}
    x_0+t_0=t_1+g(t_1),
\end{equation}
and we may solve for $t_1$.
If the intersection does not occur in the static region, we must determine the next reflection point, this time with the left mirror. Again, as $v=x-t$ is constant along right-moving light rays, we may obtain $t_2$ from this property since
\begin{equation}
    t_1-g(t_1)=t_2-f(t_2).
\end{equation}
The process is repeated until the static region is reached. Then, we may reconstruct the function $G(z)$ in the following way:
\begin{enumerate}
    \item Let $z=x_0+t_0$. Then, $G(z)=G(t_1+g(t_1))$.
    \item From the second one of Moore's equations, we have $G(t_1+g(t_1))=F(t_1-g(t_1))+2$ together with  $F(t_1-g(t_1))+2=F(t_2-f(t_2))+2$.
    \item From the first Moore's equation we have ${G(z)=F(t_2-f(t_2))+2}$, as well as condition ${F(t_2-f(t_2))+2=G(t_2+f(t_2))+2}$. 
    \item Define $z=f(t_2)+t_2$ and repeat the process until a null line
    reflecting off the right mirror enters the $F's$
    static region $[-\infty,0]$ (case 1); or a null line reflecting
    off the left one enters the $G's$ static region $[-\infty,\Lambda]$ (case 2).
    \item Every time there is a reflection off the right mirror,
    the value of $G(z)$ increases by $2$. Thus $G(z)$ will be
    two times the number of reflections off the right mirror
    plus the value of $G$ evaluated in the region $[-\infty,\Lambda]$
    or that of $F$ in $[-\infty,0]$:
    \begin{equation}
        G(z)=2n+\frac{t_f}{\Lambda}, 
    \end{equation}
    \begin{equation}\nonumber
    t_f=\left\{\mqty{z-2\qty[\sum_{i=1}^{n} g(t_{2i-1})-\sum_{i=0}^{n-1} f(t_{2i}) ],&({\rm case\,\, 1}),\\\\z-2\qty[\sum_{i=1}^{n} g(t_{2i-1})-\sum_{i=0}^{n} f(t_{2i}) ],&({\rm case\,\, 2}).}\right.
    \end{equation}
\end{enumerate}
The main limitations of this method appear when the mirrors undergo sharp accelerations or when the mirrors reach speeds close to the speed of light. In these cases, we found numerical difficulties finding the intersections between light rays and the mirrors. We hence loss considerable precision. This is also the case if one tries to explore the ultraviolet limit of the model. Besides, we can only apply this method to a massless field in (1+1)-dimensions. In other words, we do not know how to extend it to the massive or higher-dimensional cases.  The advantage of this method, on the other hand, relies in the simplicity. Our numerical simulations show that it is much faster than the finite differences method discussed in the main text. Besides, this method is very useful for analytical or semi-analytical calculations that involve the functions $G$ and $F$, and its derivatives, as long as they are of low order.

Once the expressions for the functions $F$ and $G$ have been obtained, we are able to calculate any observable of the system. For this purpose, it is very useful to give the Bogoliubov coefficients.
Indeed, let us start with a static pair of mirrors separated a distance $\Lambda_i$, and at $t\leq t_i=0$ both mirrors start to move with trajectories $f(t)$ and $g(t)$. At this time, there is a well defined basis of initial states ${\bf u}^{I}_{i}(x,t)=\delta^I_n\psi^{(i)}_n(x,t)$, with
\begin{align}
   \psi^{(i)}_n(x,t)&=\frac{1}{\sqrt{\pi n}}\sin(\tfrac{n\pi}{\Lambda_{i}}x)e^{-i\tfrac{n\pi}{\Lambda_{i}}t}=\\
   &=\frac{i}{\sqrt{4\pi n}}\qty(e^{-i\tfrac{n\pi}{\Lambda_{i}}(t+x)}-e^{-i\tfrac{n\pi}{\Lambda_{i}}(t-x)}).\notag
\end{align}

On the other hand, at times $t\geq t_f=T$, both mirrors come to rest at their final positions, $f(t)=0$ and $g(t)=\Lambda_f$. For $t\in [T,\infty)$, there is a well defined vacuum state, associated to a set of modes given by ${\bf u}^{I}_{f}(x,t)=\delta^I_n\psi^{(f)}_n(x,t)$, with
\begin{align}
   \psi^{(f)}_n(x,t)&=\frac{1}{\sqrt{\pi n}}\sin(\tfrac{n\pi}{\Lambda_{f}}x)e^{-i\tfrac{n\pi}{\Lambda_{f}}t}=\\
   &=\frac{i}{\sqrt{4\pi n}}\qty(e^{-i\tfrac{n\pi}{\Lambda_{f}}(t+x)}-e^{-i\tfrac{n\pi}{\Lambda_{f}}(t-x)}).\notag
\end{align}
As we have seen, in the interval $[0,T]$, the elements of the basis ${\bf u}^{I}_{i}(x,t)$ evolve nontrivially due to the time-dependent boundary conditions. The evolved modes are written in terms of the functions $G(z)$ and $F(z)$ as in Eq. \eqref{eq:evomodes}. The nontrivial evolution of these field modes implies that the initial vacuum state may also evolve into a state which differs from the vacuum state in the final, static region. In that case, we say that there have been a particle creation process. The number of particles created in the final state will be a function of the Bogoliubov coefficients $\alpha$ and $\beta$ that relate the original, evolved modes to the final vacuum modes. If we express
\begin{equation}
    \psi^{(i)}_m = \sum_n \alpha_{nm} \psi^{(f)}_n-\beta_{nm} \bar\psi^{(f)}_n,
\end{equation}
these Bogoliubov coefficients are defined by means of the Klein-Gordon products $\alpha_{nm}= \langle\psi^{(f)}_n,\psi^{(i)}_m\rangle$ and $\beta_{nm}=\langle\bar \psi^{(f)}_n,\psi^{(i)}_m\rangle$, namely,
\begin{align}\nonumber
    \alpha_{nm}&=-i\int_{f(t)}^{g(t)}\qty(\bar\psi^{(f)}_n\partial_t\psi^{(i)}_m-\psi^{(i)}_m\partial_t\bar\psi^{(f)}_n)dx,\\
    \beta_{nm}&=-i\int_{f(t)}^{g(t)}\qty(\psi^{(f)}_n\partial_t\psi^{(i)}_m-\psi^{(i)}_m\partial_t\psi^{(f)}_n)dx.
\label{eq:alphabetas}
\end{align}
The integrals in \eqref{eq:alphabetas} are time independent (up to an irrelevant phase)\footnote{Due to the fact that the Klein Gordon product between solutions is time independent}, so we may choose to calculate them at $t=T$. Let us define $\tilde{\psi}_n=i(e^{-in\pi G}+e^{-in\pi F})/\sqrt{4\pi n}$. Then, we have $\partial_t\psi^{(f)}_n=\partial_x\tilde{\psi}_n$ and write

\begin{widetext}
\begin{align}
        \alpha_{nm}&=-i\int_{f(t)}^{g(t)}\qty(\bar\psi^{(f)}_n\partial_t\psi^{(i)}_m-\psi^{(i)}_m\partial_x\bar{\tilde{\psi}}_n)dx=-i\int_{f(t)}^{g(t)}\qty(\bar\psi_n^{(f)}\partial_t\psi^{(i)}_m+\partial_x\psi^{(i)}_m\bar{\tilde{\psi}}_n)dx \notag \\[2mm]
        &=\frac{1}{2\Lambda_f}\sqrt{\frac{m}{n}}\int _0^{\Lambda_f}
\qty{e^{-i\pi\qty[m\tfrac{(t+x)}{\Lambda_f} -nG(t+x)]}+e^{-i\pi\qty[m\tfrac{(t-x)}{\Lambda_f} -nF(t-x)]}}dx,    
\end{align}
\begin{align}
        \beta_{nm}&=-i\int_{f(t)}^{g(t)}\qty(\psi^{(f)}_n\partial_t\psi^{(i)}_m-\psi^{(i)}_m\partial_x\tilde{\psi}_n)dx=-i\int_{f(t)}^{g(t)}\qty(\psi_n^{(f)}\partial_t\psi^{(i)}_m+\partial_x\psi^{(i)}_m\tilde{\psi}_n)dx \notag \\[2mm]
        &=\frac{1}{2\Lambda_f}\sqrt{\frac{m}{n}}\int _0^{\Lambda_f}
\qty{e^{-i\pi\qty[m\tfrac{(t+x)}{\Lambda_f} +nG(t+x)]}+e^{-i\pi\qty[m\tfrac{(t-x)}{\Lambda_f} +nF(t-x)]}}dx,    
\end{align}
\noindent
where we have integrated by parts and used the boundary conditions. Defining the new variables $u=(x+t)/\Lambda_f$ and $v=(t-x)/\Lambda_f$, we may rewrite these expressions as
\begin{align}
    \alpha_{nm}&=\frac{1}{2}\sqrt{\frac{m}{n}}\Bigg\{\int _{t/\Lambda_f}^{t/\Lambda_f+1}
e^{-i\pi\qty[mu -nG(\Lambda_f u)]}du+\int _{t/\Lambda_f-1}^{t/\Lambda_f}e^{-i\pi\qty[mv -nF(\Lambda_f v)]}dv\Bigg\},
\end{align}
\begin{align}
    \beta_{nm}&=\frac{1}{2}\sqrt{\frac{m}{n}}\Bigg\{\int _{t/\Lambda_f}^{t/\Lambda_f+1}
e^{-i\pi\qty[mu +nG(\Lambda_f u)]}du+\int _{t/\Lambda_f-1}^{t/\Lambda_f}e^{-i\pi\qty[mv +nF(\Lambda_f v)]}dv\Bigg\}.
\end{align}
\end{widetext}
We should keep in mind that these expressions are valid only for $t\geq T$, namely, at times where both mirrors come to rest at their final positions, $f(t)=0$ and $g(t)=\Lambda_f$. Besides, the coefficients will be trivial if and only if $G(u)=u$ and $F(v)=v$ for all $u$ and $v$. 

We have checked that the Bogoliubov coefficients obtained via the conformal transformation method agree with those obtained solving Hamilton's equations up to a certain numerical accuracy for some simple trajectories. Here, we show the first confirguration discussed in Sec. \ref{sec:2-mov-boundar}. This is illustrated in Fig. \ref{fig:betas_comparison} by comparing the (squared) Bogoliubov $\beta$ coefficients obtained from both methods for a trajectory similar to Eq. \eqref{eq:fg-traj}, but with the switching function $B(t)$ being a simple Gaussian (instead of compactly supported bump function that gives rise to some numerical issues for the conformal method).
\begin{figure}
    \centering
\includegraphics[width = 0.49\textwidth]{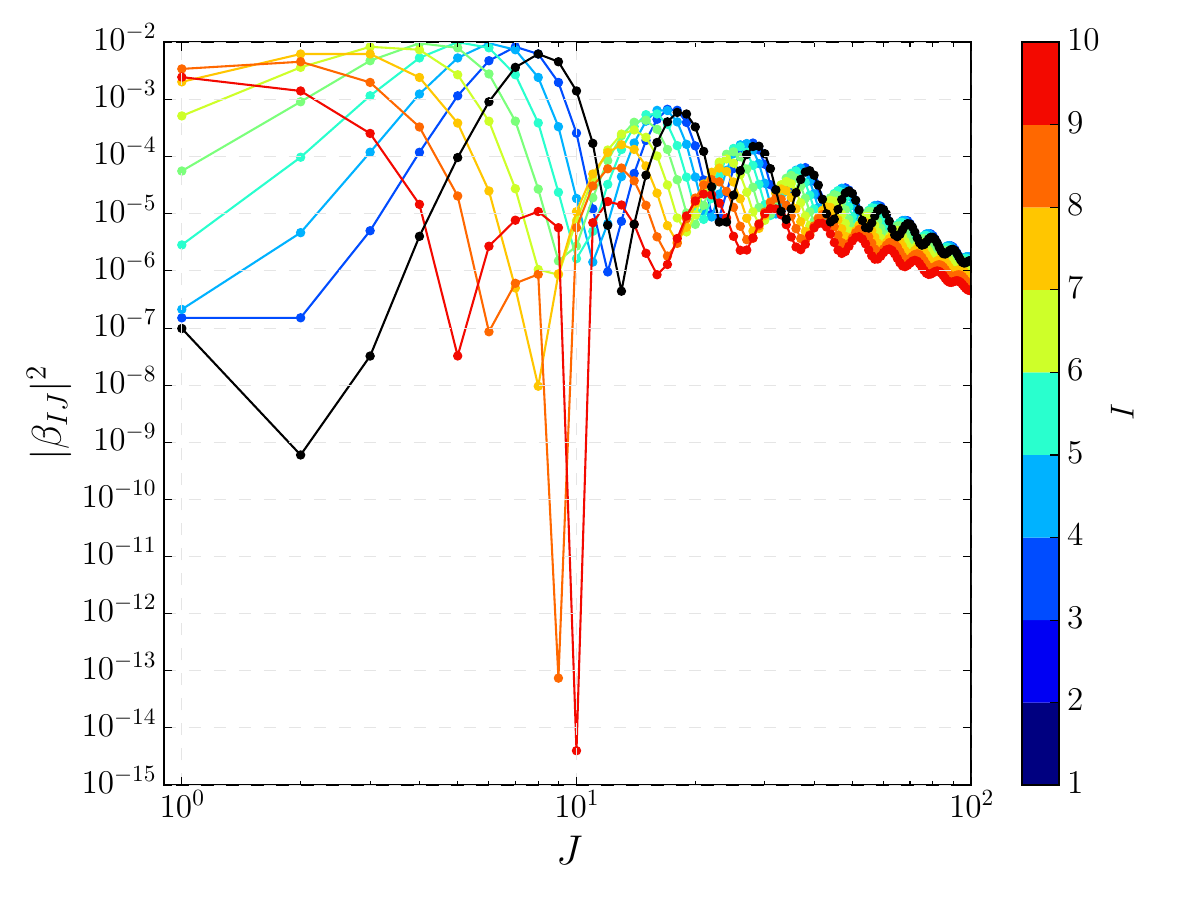}
\includegraphics[width = 0.49\textwidth]{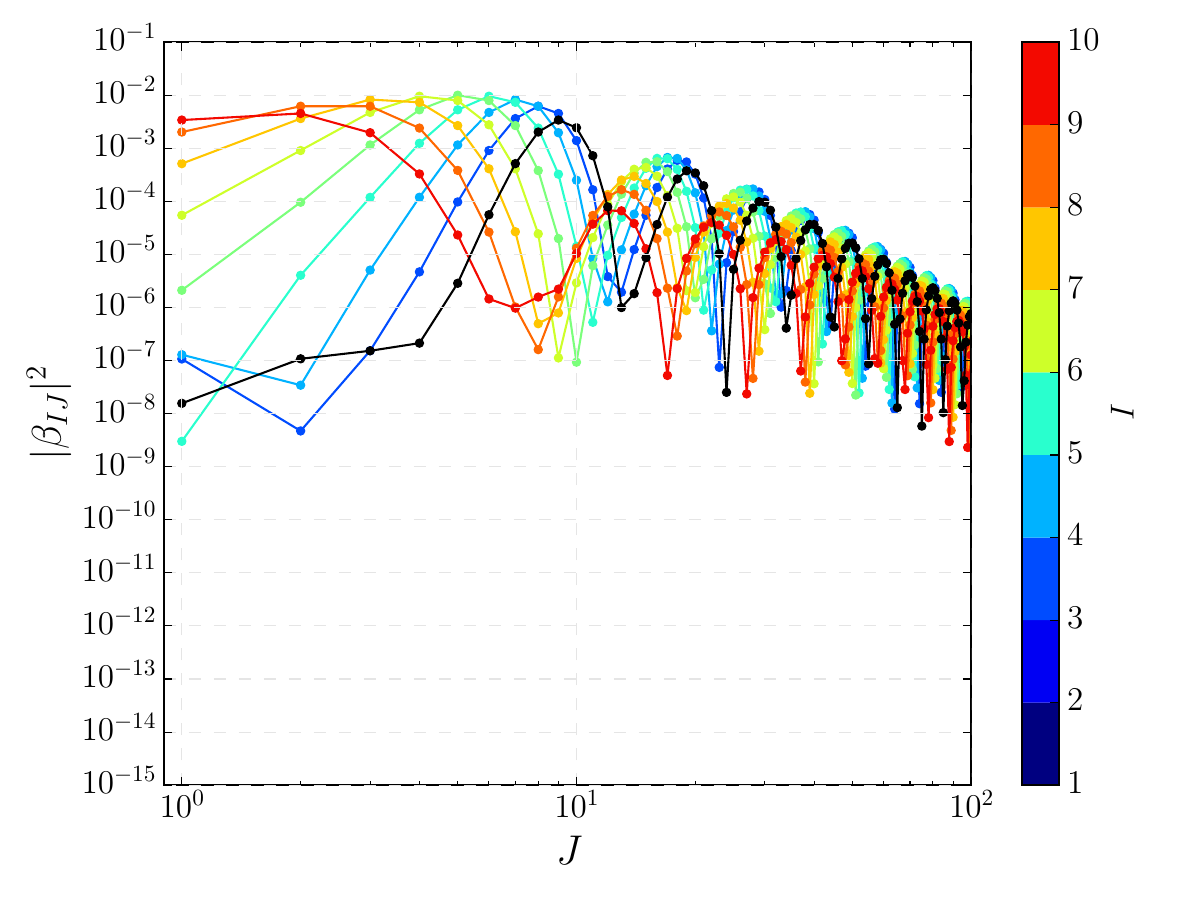}
    \caption{Comparison of the spectral density $|\beta_{IJ}|^2$ as obtained from the conformal (upper panel) and canonical (lower panel) methods.}
    \label{fig:betas_comparison}
\end{figure}
An estimation of the numerical error of the conformal method can be obtained by computing how well Moore's equations \eqref{Moores1} and \eqref{Moores2} are satisfied throughout the evolution. We plot such conditions in Fig. \eqref{fig:Gz_errors}. The resulting Moore's functions $F(z)$ and $G(z)$, with $F(z)=G(z)$ for this particular configuration of the boundaries, present the characteristic stairlike shape during the period of nontrivial acceleration of the mirror, and again becomes trivial after the mirror reaches the static phase.
\begin{figure}
    \includegraphics[scale=0.6]{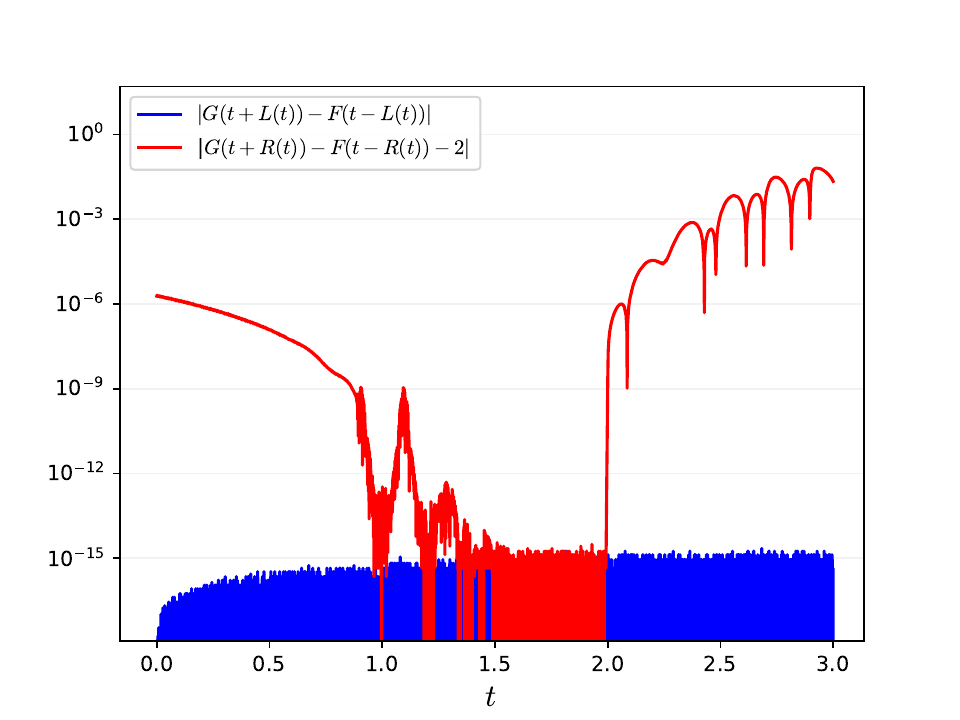}
    \includegraphics[scale=0.55]{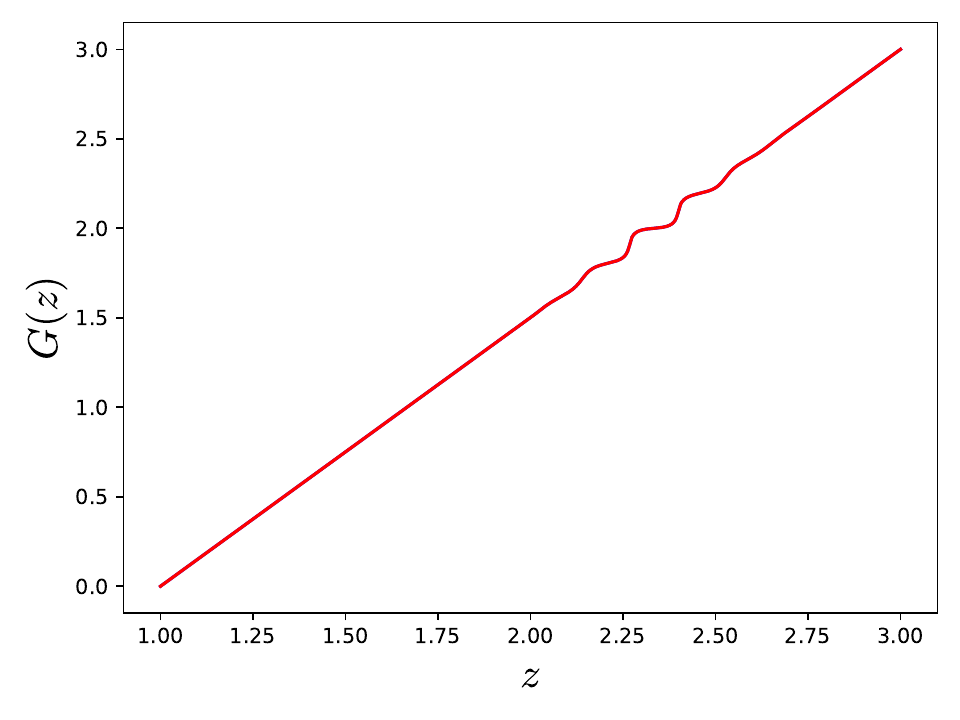}
    \caption{Accuracy of Moore's equations throughout the evolution of the quantum field during the relevant period in which one mirror is moving (upper panel) and resulting Moore function (lower panel).}
    \label{fig:Gz_errors}
\end{figure}
Although  we have restricted our analysis to the study of observables that do not require renormalization, the conformal method is well-suited to study semi-analytically some of the observables that require renormalization, for instance the components of the renormalized stress-energy tensor (RSET). Indeed, following Fulling and Davies \cite{Davies1976}, we can write the components of the RSET as
\begin{align}
\left\langle T_{00}\right\rangle&=-[f_G(t+x)+f_F(t-x)]=\left\langle T_{11}\right\rangle, \\[2mm]
\left\langle T_{01}\right\rangle&=f_G(t+x)-f_F(t-x)=\left\langle T_{10}\right\rangle ,
\end{align}
with
\begin{align}
    f_G&=\frac{1}{24 \pi}\left[\frac{G^{\prime \prime \prime}}{G^{\prime}}-\frac{3}{2}\left(\frac{G^{\prime \prime}}{G^{\prime}}\right)^2+\frac{1}{2} \pi^2\left(G^{\prime}\right)^2\right],\\
     f_F&=\frac{1}{24 \pi}\left[\frac{F^{\prime \prime \prime}}{F^{\prime}}-\frac{3}{2}\left(\frac{F^{\prime \prime}}{F^{\prime}}\right)^2+\frac{1}{2} \pi^2\left(F^{\prime}\right)^2\right].
\end{align}
In Fig.~\ref{fig:T00_traj} we plot the renormalized energy density $\langle T_{00}(t,x=g(t))\rangle$ as a function of time. We clearly see that the energy flux is peaked in the regions of large acceleration of the mirror.
\begin{figure}
    \centering
    \includegraphics[scale=0.5]{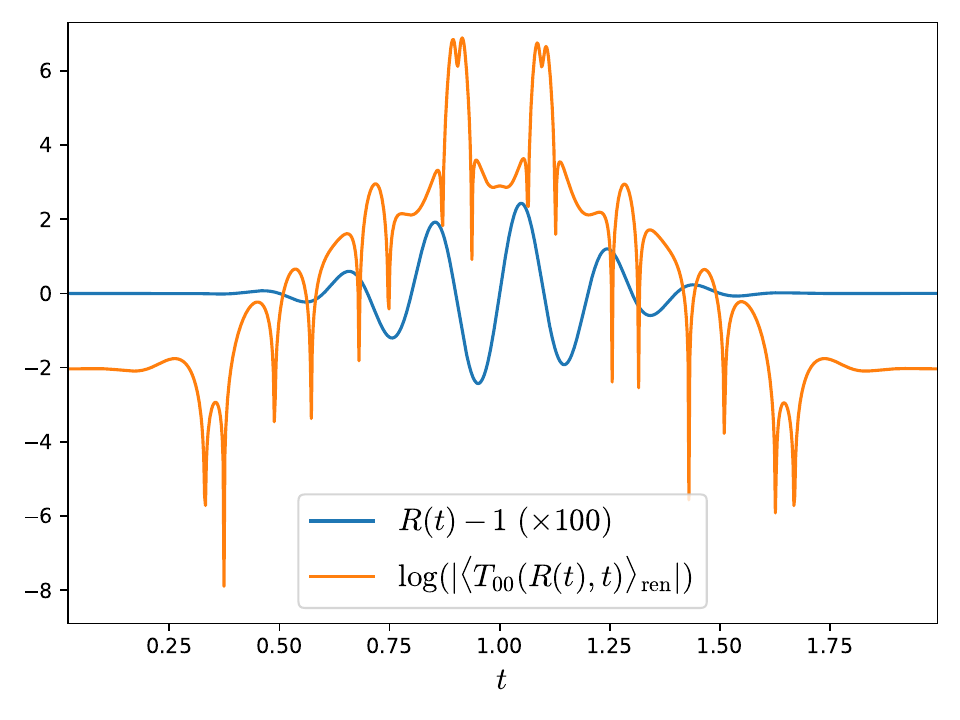}
    \caption{Renormalized energy density at the right boundary $x=g(t)$ and trajectory as a function of time.}
    \label{fig:T00_traj}
\end{figure}

\section{Massive 3+1 theory}
\label{Subsec:3+1}
Our analysis and codes can be straightforwardly extended to the $3+1$ massive scalar field theory inside a box with the shape of a parallelepiped, see Fig.~\ref{fig:3d} for a representation. More general geometries can also be considered since they are relevant for some of the current experiments and constitute ongoing work. Among the six boundaries that our system displays, just two opposite boundaries of it will move. We choose those boundaries to be orthogonal to the $\hat x$ direction, while the boundaries orthogonal to the directions $\hat y$ and $\hat z$ are fixed. 
\begin{figure}
{\centering     
  \includegraphics[width = 0.35\textwidth]{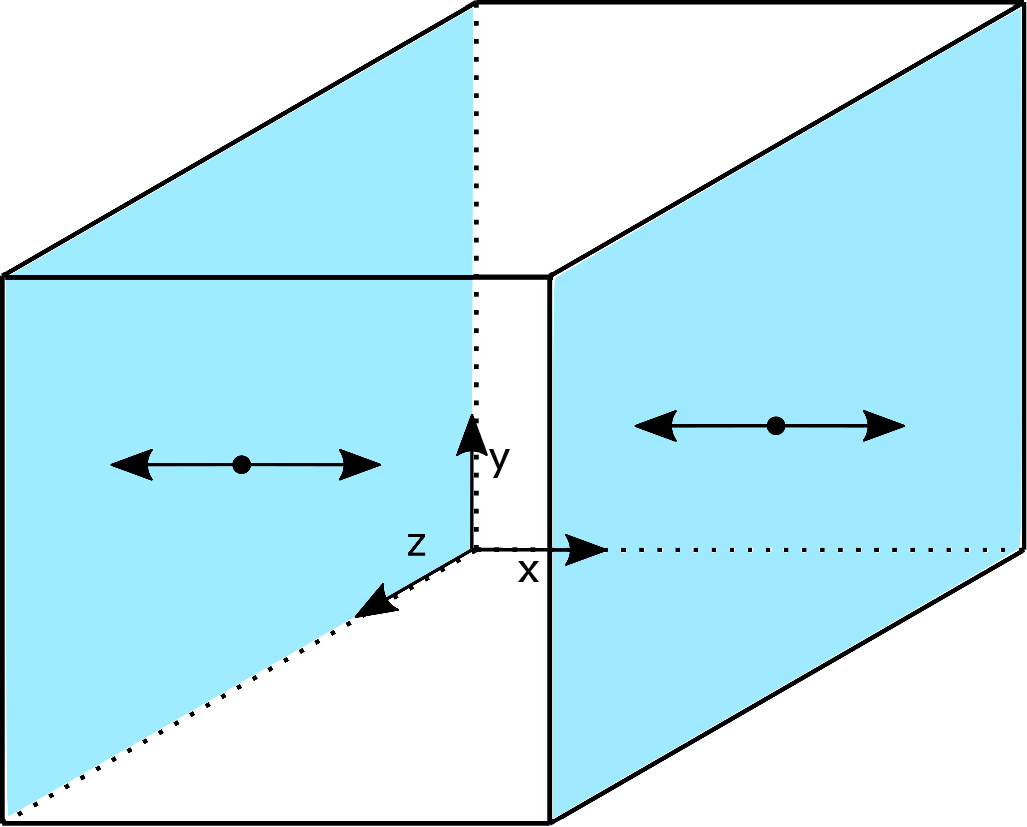}
}
\caption{Pictorial representation of the setup we explain in the text. Shaded in blue we display the two moving boundaries, corresponding to the boundaries that have $\hat x$ as their normal direction.}
\label{fig:3d}
\end{figure}

The action that we will study now is
\begin{equation}\label{eq:action-2}
    S=- \frac{1}{2} \int dt \int d^3x \sqrt{-\eta} \left( \eta^{\mu \nu} \partial_{\mu} \phi \partial_{\nu} \phi + m^2\phi^2\right),
\end{equation}
leading to the equations of motion
\begin{align}
    \frac{1}{\sqrt{-\eta}} \partial_{\mu} \left( \sqrt{-\eta} \eta^{\mu \nu} \partial_{\nu} \right) \phi - m^2 \phi= 0. 
\end{align}
The range of the coordinates now is $t \in \mathbb{R}$, $x \in (f(t),g(t))$ for a given $t$, $y \in (0,L_y)$ and $z \in (0, L_z)$. Now, we have the following boundary conditions for the field: $ \phi\left(t,f(t),y,z \right) = \phi (t,g(t),y,z) = \phi(t,x,0,z) = \phi(t,x,L_y,z) = \phi (t,x,y,0) = \phi(t,x,y,L_z) = 0$.  
Therefore, we still can apply the coordinate transformation in Eq. \eqref{eq:coordtrans} in the $(t,x)$ sector and move to new coordinates $(\tau,\xi,y,z)$. As in the 1+1 case described above, we assume that $f(t)=0$ and $g(t)=L_0$ when $t\to-\infty$. Hence, we will again measure everything in units of $L_0$, putting it to one. Repeating the same change of coordinates that we introduced for the $1+1$-dimensional setup in the $(t,x)$ coordinates we can write the metric as the product of the metric from Eq.~\eqref{eq:accoustic-g} and two flat directions
\begin{align}
ds^2=& -\left(1-V^2(\tau,\xi)\right)d\tau^2+2L(\tau)V(\tau,\xi)d\tau d\xi \nonumber  \\ 
& +L^2(\tau)d\xi^2 + dy^2 + dz^2, \label{eq:accoustic-3d}
\end{align}
Now the Hamiltonian contains some additional terms associated with the additional flat directions and the non-zero mass.  
\begin{eqnarray}\nonumber
&&H_{T}=\int_{\cal V}d{\cal V}\bigg[\frac{N}{2} \frac{\pi_{\phi}^{2}}{L}+\frac{N}{2 L}\left(\partial_\xi\phi\right)^{2}+\left(\xi \frac{\dot L}{L}+\frac{\dot f}{L}\right) (\partial_\xi\phi)\pi_{\phi}\\\label{eq:3dham}
&&+\left(\partial_y\phi\right)^{2}+\left(\partial_z\phi\right)^{2}+\frac{1}{2}m^2\phi^2\bigg],
\end{eqnarray}
where we have introduced the notation
\begin{equation}
\int_{\cal V}d{\cal V}=\int_{0}^{L_z}dz\int_0^{L_y}dy\int_{0}^{1} d \xi.
\end{equation}
They satisfy the Poisson brackets at equal times
\begin{equation}\label{eq:position-poisson3D}
\left\{\phi(\tau,\xi,y,z), \pi_{\phi}(\tau,\tilde{\xi},\tilde{y},\tilde{z})\right\}=\delta(\xi-\tilde{\xi})\delta(y-\tilde{y})\delta(z-\tilde{z}).
\end{equation}
If we expand in normal modes, 
\begin{eqnarray}
 \phi(\tau,\xi,y,z) = & \sum_{\bf n} \phi_{\bf n} (\tau)\sin (n_1 \pi \xi) \nonumber \\ 
 & \times \sin \left(\frac{n_2 \pi y}{L_y}\right)\sin \left(\frac{n_3 \pi z}{L_z}\right), \nonumber \\ 
 \pi_\phi(\tau,\xi,y,z) = & \sum_{\bf n} (\tau)\pi_{\bf n}\sin (n_1 \pi \xi) \nonumber \\
 & \times \sin \left(\frac{n_2 \pi y}{L_y}\right)\sin \left(\frac{n_3 \pi z}{L_z}\right)\label{eq:fourier-3D},
\end{eqnarray}
where ${\bf n}=(n_1,n_2,n_3)$, with $n_i\in\mathbb{N}^+$.  This Fourier transform and the canonical Poisson brackets amount to 
\begin{equation}\label{eq:fourier-poisson3d}
\left\{\phi_{\bf n} (\tau), \pi_{{\bf n'}} (\tau) \right\}=2 \delta_{{\bf n n'}}.
\end{equation}
One can easily write the Hamiltonian \eqref{eq:3dham} in terms of the Fourier modes, and compute the Hamilton's equations of motion. In this case we get 
\begin{align}
 &\dot \phi_{\bf n}=\frac{1}{L} \pi_{\bf n}-\frac{\dot L}{2 L} \phi_{\bf n} +2 \sum_{m_1}\left(1-\delta_{m_1 n_1}\right)\frac{m_1 n_1}{m_1^{2}-n_1^{2}}\nonumber \\\label{eq:dotphi3d}
& \times\left[\frac{\dot f}{L}\left((-1)^{m_1+n_1}-1\right)+\frac{\dot L}{L}(-1)^{m_1+n_1}\right]  \phi_{(m_1,n_2,n_3)}\\ \nonumber
&\dot{\pi}_{\bf n}=-\frac{1}{L}\left((n_1 \pi)^{2} + \frac{(n_2 \pi)^{2}}{ L_y^2}+\frac{(n_3 \pi)^{2}}{ L_z^2}+m^2\right) \phi_{\bf n}\\ \nonumber
&+\frac{\dot L}{2 L} \pi_{\bf n} + 2 \sum_{m_1}\left(1-\delta_{n_1 m_1}\right)\frac{n_1 m_1}{m_1^{2}-n_1^{2}}\nonumber \\\label{eq:dotpi3d}
& \times\left[\frac{\dot f}{L}\left((-1)^{n_1+m_1}-1\right)+\frac{\dot L}{L}(-1)^{n_1+m_1}\right]  \pi_{(m_1,n_2,n_3)}.
\end{align}
It is interesting to note that there are not any modes with vanishing momentum on the perpendicular directions to the moving boundary, and hence all the created particles carry momentum on those transverse directions.

The basis of orthonormal complex solutions to the above equations of motion will be denoted by ${\bf u}^{(\bf I)}(\tau)$, with ${\bf I}=(I_1,I_2,I_3)$. The inner product between two elements of the basis will be now given by 
\begin{align}\label{eq:inner-prod3d}
&\langle {\bf u}^{(\bf I)}(\tau), {\bf u}^{(\bf J)}(\tau)\rangle=\frac{i}{2} \sum_{\bf n} \bar{\phi}_{\bf n}^{(\bf I)}(\tau) \pi_{\bf n}^{(\bf J)}(\tau)-\bar{\pi}_{\bf n}^{(\bf I)}(\tau) \phi_{\bf n}^{(\bf J)}(\tau).
\end{align}
Since the solutions are orthonormal, and given the properties of the inner product, they will again satisfy $\langle {\bf u}^{(\bf I)}(\tau), {\bf u}^{(\bf J)}(\tau)\rangle=\delta^{\bf IJ}$, $\langle {\bf u}^{(\bf I)}(\tau), \bar{\bf u}^{(\bf J)}(\tau)\rangle=0$ and $\langle \bar{\bf u}^{(\bf I)}(\tau), \bar{\bf u}^{(\bf J)}(\tau)\rangle=-\delta^{\bf IJ}$. The remain of the canonical analysis is completely parallel to the one we have introduced in~\ref{Sec:classical}. One just needs to replace the indexes $(I,J,\ldots )$ by $({\bf I},{\bf J},\ldots )$ and $(n,m,\ldots)$ by $({\bf n},{\bf m},\ldots)$. We highlight that the conformal transformation method cannot be used to solve now the problem (not even on the massless case). However, the numerical solution in which the number of modes is truncated and then extrapolated to an arbitrary number of modes can still be used though the complexity clearly increases due to the growth in the number of coupled modes.

\bibliography{dce_biblio_PRD}

\end{document}